\documentclass{emulateapj}

\def\gsim{\mathrel{\rlap{\lower 4pt \hbox{\hskip 1pt $\sim$}}\raise 1pt
\hbox {$>$}}}
\def\lsim{\mathrel{\rlap{\lower 4pt \hbox{\hskip 1pt $\sim$}}\raise 1pt
\hbox {$<$}}}

\usepackage{color}

\begin{document}

\title{Sodium Absorption Systems toward SN Ia 2014J Originate on Interstellar Scales$^{20}$
}  

\author{
K. Maeda\altaffilmark{1,2}, A. Tajitsu\altaffilmark{3}, K.S. Kawabata\altaffilmark{4,5}, R.J. Foley\altaffilmark{6}, S. Honda\altaffilmark{7,8}, Y. Moritani\altaffilmark{2,4}, \\
M. Tanaka\altaffilmark{9}, O. Hashimoto\altaffilmark{8}, M. Ishigaki\altaffilmark{2}, J.D. Simon\altaffilmark{10}, M.M.Phillips\altaffilmark{11}, M. Yamanaka\altaffilmark{12,13}, \\
D. Nogami\altaffilmark{1,13}, A. Arai\altaffilmark{7}, W. Aoki\altaffilmark{9}, K. Nomoto\altaffilmark{2,19}, D. Milisavljevic\altaffilmark{14}, P.A. Mazzali\altaffilmark{15,16}, \\
A.M. Soderberg\altaffilmark{14}, M. Schramm\altaffilmark{2,17}, B. Sato\altaffilmark{18}, H. Harakawa\altaffilmark{9,18}, N. Morrell\altaffilmark{11}, N. Arimoto\altaffilmark{3}
}

\altaffiltext{20}{Based on data collected at the Subaru Telescope and Okayama 1.88m Telescope, which are operated by the National Astronomical Observatory of Japan, and at the Gunma 1.5m Telescope operated by the Gunma Astronomical Observatory.}
\altaffiltext{1}{Department of Astronomy, Kyoto University, Kitashirakawa-Oiwake-cho, Sakyo-ku, Kyoto 606-8502, Japan; keiichi.maeda@kusastro.kyoto-u.ac.jp .}
\altaffiltext{2}{Kavli Institute for the Physics and Mathematics of the 
Universe (WPI), The University of Tokyo, 5-1-5 Kashiwanoha, Kashiwa, Chiba 277-8583, Japan}
\altaffiltext{3}{Subaru Telescope, National Astronomical Observatory of Japan, 650 North A'ohoku Place, Hilo, Hawaii 96720} 
\altaffiltext{4}{Hiroshima Astrophysical Science Center, Hiroshima University, Kagamiyama, Higashi-Hiroshima, Hiroshima 739-8526, Japan} 
\altaffiltext{5}{Department of Physical Science, Hiroshima University, Kagamiyama, Higashi-Hiroshima 739-8526, Japan} 
\altaffiltext{6}{Astronomy Department, University of Illinois at Urbana-Champaign, 1002 W. Green Street, Urbana, IL 61801 / Department of Physics, University of Illinois at Urbana-Champaign, 1110 W. Green Street, Urbana, IL 61801}
\altaffiltext{7}{Nishi-Harima Astronomical Observatory, Center for Astronomy, University of Hyogo, 407-2, Nishigaichi, Sayo-cho, Sayo, Hyogo 679-5313, Japan}
\altaffiltext{8}{Gunma Astronomical Observatory, Takayama, Gunma 377-0702, Japan}
\altaffiltext{9}{National Astronomical Observatory of Japan, Mitaka, Tokyo 181-8588, Japan} 
\altaffiltext{10}{Observatories of the Carnegie Institution for Science, 813 Santa Barbara St, Pasadena, CA 91101}
\altaffiltext{11}{Carnegie Observatories, Las Campanas Observatory, Casilla 601, La Serena, Chile}
\altaffiltext{12}{Department of Physics, Faculty of Science and Engineering, Konan University, Okamoto, Kobe, Hyogo 658-8501, Japan}
\altaffiltext{13}{Kwasan Observatory, Kyoto University, 17-1 Kitakazanohmine-cho, Yamashina-ku, Kyoto, 607-8471}
\altaffiltext{14}{Harvard-Smithsonian Center for Astrophysics, 60 Garden Street, Cambridge, MA 02138}
\altaffiltext{15}{Astrophysics Research Institute, Liverpool John Moores University, Liverpool L3 5RF, UK}
\altaffiltext{16}{Max-Planck-Institut f\"ur Astrophysik, Karl-Schwarzschild-Strasse 1, D-85748 Garching, Germany}
\altaffiltext{17}{Frequency Measurement Group, National Institute of Advanced Industrial Science and Technology (AIST) Tsukuba-central 3-1, Umezono 1-1-1, Tsukuba, Ibaraki 305-8563, Japan} 
\altaffiltext{18}{Department of Earth and Planetary Sciences, Tokyo Institute of Technology, 2-12-1 Ookayama, Meguro-ku, Tokyo 152-8551, Japan} 
\altaffiltext{19}{Hamamatsu Professor}

\begin{abstract}
  \ion{Na}{1}~D absorbing systems toward Type Ia supernovae (SNe~Ia)
  have been intensively studied over the last decade with the aim of
  finding circumstellar material (CSM), which is an indirect probe of
  the progenitor system. However, it is difficult to deconvolve
    CSM components from non-variable, and often dominant, components
    created by interstellar material (ISM). We present a series of
  high-resolution spectra of SN~Ia~2014J from before maximum
  brightness to $\gsim$250~days after maximum brightness. The
  late-time spectrum provides unique information for determining the
  origin of the \ion{Na}{1}~D absorption systems. The deep late-time
  observation allows us to probe the environment around the SN at a
  large scale, extending to $\gsim$40~pc. We find that a spectrum of
  diffuse light in the vicinity, but not directly in the line-of-sight, 
  of the SN has absorbing systems nearly identical to those obtained
  for the `pure' SN line-of-sight.  Therefore, basically all
  \ion{Na}{1}~D systems seen toward SN~2014J must originate from
  foreground material that extends to at least $\sim$40 pc in
  projection and none at the CSM scale. A fluctuation in the column
  densities at a scale of $\sim$20~pc is also identified.  After
  subtracting the diffuse, ``background'' spectrum, the late-time SN
  \ion{Na}{1}~D profile along the SN line-of-sight is consistent with the profile 
 in the near-maximum brightness spectra.  The lack of variability on a
  $\sim$1~year timescale is consistent with the ISM interpretation for
  the gas.
\end{abstract}


\section{Introduction}

Despite their importance for cosmology as standardizable candles and 
for astrophysics as an origin of Fe-peak elements, the progenitor system
of Type Ia supernovae (SNe~Ia) has remained as an unresolved and
extensively discussed question for many decades \citep[see, e.g.,][for
a review]{hillebrandt2000}. In particular, there have been ongoing
efforts to discriminate between the so-called single-degenerate (SD)
scenario and the double-degenerate (DD) scenario: in the former an
SN~Ia is a thermonuclear explosion of a nearly Chandrasekhar-mass
white dwarf (WD) formed by the accretion of material from a
non-degenerate companion star \citep[e.g.,][]{whelan1973, nomoto1982,
  hachisu1999}, while in the latter an SN~Ia comes from the merger of
two WDs \citep[e.g., ][]{iben1984, webbink1984}.

There are various observational features proposed so far to
  diagnose a progenitor system of an individual SN as either SD or DD.
  These methods include the detection of excess emission in the {\em
    UV} and the blue portion of the optical region in the first few days
  after a SN~Ia explosion.  The early excessive emission is suggested
  to be a signature of the impact between the SN ejecta and a
  non-degenerate binary companion in the SD scenario \citep{kasen2010,
    kutsuna2015}.  Detections of this behavior have been reported for the
  peculiar SN~2002es-like iPTF14atg \citep{cao2015} and the normal
  SN~Ia~2012cg \citep{marion2015b}. Among other diagnostics, the
environment just around the SN progenitor is key, since the SD
scenario predicts a relatively `dirty' environment (i.e., having
significant circumstellar material; `CSM') through mass loss from the
companion and/or nova eruptions, while the DD scenario is thought to
lead to a relatively `clean' environment (i.e., a circumstellar
environment similar to or less dense than the interstellar medium;
`ISM').  This picture is complicated by the possibility that a DD
progenitor system may also produce a dense CS environment in some
cases \citep[e.g.,][]{shen2013, raskin2013, tanikawa2015}, but in any
case the CSM environment provides strong constraints on and insight into 
progenitor evolution. There is also another scenario, the core-degenerate (CD) model (Sparks \& Stecher 1974), which could produce even more massive CSM than the SD \citep{soker2015}. We refer to \citet{maoz2014} for a review of observational constraints on the progenitor systems of SNe~Ia.

Among various methods proposed to probe the environment and existence
of CSM, increasing attention has been given to
narrow-line absorption systems in the lines-of-sight toward SNe~Ia over the last decade.
SNe offer a unique opportunity to probe diffuse gas in galaxies beyond
the local group.  Material along the line of sight is backlit by the
SN and is detected as absorption features in (especially
high-resolution) spectra \citep[e.g.,][]{rich1987,cox2008}.  However,
distinguishing CSM and ISM components is challenging.  The SN produces
a strong UV radiation field, which can ionize some atomic species if
they are sufficiently close.  The ionized atoms will then recombine on
a timescale of weeks to months, producing variable absorption on this
timescale.  Therefore, time-variable absorption features provide
strong evidence of CSM \citep{patat2007,simon2009,sternberg2014}.
However, the lack of variability does not imply a lack of CSM, and it
has not been clarified for any individual SNe~Ia if there are `hidden'
CSM components in the non-variable absorption systems. For a
statistical sample of SNe~Ia, an imbalance between \ion{Na}{1}~D
systems with ``blueshifted'' and ``redshifted'' profiles, with there
being more ``blueshifted'' SNe, is indirect evidence for outflows
expected to be associated with CSM \citep{sternberg2011, foley2012,
  maguire2013}.  This fact has been used as evidence for a population
that evolves through the SD path, highlighting the importance of
investigating the existence of hidden CSM components in the non-variable
\ion{Na}{1}~D systems.

SN~Ia~2014J in M82 ($d \approx 3.8$~Mpc), discovered by
\citet{fossey2014}, is the closest SN~Ia in the last few decades
and provides a unique opportunity to investigate the progenitor issue.
Thanks to its proximity, intensive follow-up observations at
  various wavelengths have been performed for SN~2014J. Optical and
  NIR observations show that SN~2014J belongs to a class of normal
  SNe~Ia with layered abundance stratification, but with relatively
  high velocities of absorption lines \citep[e.g.,][]{goobar2014,
    zheng2014, marion2015a, vacca2015}, in accordance with a spectral
  synthesis model \citep{ashall2014}. At late times, there is no
  detection of H$\alpha$ emission to deep limits, which has been
  suggested to be a signature of a non-degenerate companion star 
\citep{lundqvist2015}. Mid-IR
  follow-up shows possible enhancement of stable Ni in the inner part
  of the ejecta \citep{telesco2015}, which favors an explosion of a
  Chandrasekhar-mass WD. SN~2014J also became the first (and so far
  only) example of an SN~Ia for which MeV emissions from radioactive
  decays of $^{56}$Co were solidly detected \citep{churazov2014}; there was also
  probable detection of $^{56}$Ni decay emissions, which might
  favor an explosion of a sub-Chandrasekhar mass WD with a He
  companion star \citep{diehl2014}. Pre-explosion {\it Hubble Space Telescope (HST)}/{\it
    Chandra} images show no apparent source at the position of the SN,
  excluding a red-giant companion star \citep{kelly2014} and a
  supersoft X-ray source \citep{nielsen2014}, which is inconsistent with some
  variants within the SD model. There are also various observations
  wthat probe the CSM around SN~2014J: there are non-detections of
  radio and X-ray emission from SN~2014J, which limits the amount of
  CSM within $\sim$0.01~pc of the SN progenitor \citep{perez2014,
    margutti2014}.  Analyses of the extinction curve toward SN~2014J
  including {\em UV} bands result in mixed interpretations --- it has
  been agreed that the extinction curve toward SN~2014J is
  characterized by a low value of $R_{V}$, while the extinction law is
  reported to consist of either a single component \citep{brown2015,amanullah2015} or multiple-components \citep{foley2014, hoang2015}.
  If attributed to a single dust component within the ISM, it requires
  small grains \citep{gao2015, hoang2015}, which is also consistent
  with a non-standard polarization curve toward SN~2014J
  \citep{kawabata2014, patat2014}. In summary, there are multiple and
  different indications of the progenitor system of SN~2014J, some
  favoring the SD but others the DD model. Therefore, adding new and independent
  information is highly valuable. 

SN~2014J was closer than another nearby SN~Ia, 2011fe in M101 ($d
\approx 6.4$~Mpc) which unlike SN~2014J did not have significant
extinction.  SN~2014J is thus an ideal target with which one can study
the origin of the absorption systems with high-spectral resolution
observations. Observations by \citet{goobar2014} showed a strong and
complex structure of the \ion{Na}{1}~D absorption systems toward
SN~2014J, as expected from the red color of the SN itself.
\citet{welty2014} presented spectra showing strong absorption systems
not only in \ion{Na}{1}~D, but also in \ion{K}{1}, \ion{Ca}{1},
\ion{Ca}{2}, CH, CH$^{+}$, CN, and a number of diffuse interstellar
bands (DIBs), mainly seen in the wavelength ranges corresponding to
the saturated component(s) of the \ion{Na}{1}~D systems \citep[see also][]{jack2015}.  \citet{foley2014} reported no variability in \ion{Na}{1}~D and \ion{K}{1} beyond the noise level of
their spectra in the period from $-9.6$~days to $+18.4$~days, further
confirmed by \citet{ritchey2015} in their spectra taken between $-5.6$
and $+30.4$~days. \citet{graham2014} also found no variability in
\ion{Na}{1}~D between $-11$ and $+22$~days, but claimed detection of
variability in \ion{K}{1} for the most blueshifted components at
$-144$~km~s$^{-1}$ and possibly at $-127$~km~s$^{-1}$, which
correspond to an unsaturated component and a saturated one in the
\ion{Na}{1}~D absorption systems, respectively. 
If this K I variability is caused by the CSM, it would require a few $M_{\odot}$ at the SN vicinity. This led \citet{soker2015} to conclude that this is only explained by the CD scenario if the absorption systems indeed originate on the CS scale

\begin{deluxetable*}{ccccccccl}
 \tabletypesize{\scriptsize}
 \tablecaption{Log of Spectroscopic Observations
 \label{tab1}}
 \tablewidth{0pt}
 \tablehead{
   \colhead{MJD}
 & \colhead{Phase}
 & \colhead{Telescope\tablenotemark{a}}
 & \colhead{Instrument}
 & \colhead{Resolution}
 & \colhead{Wavelength Range (\AA)}
 & \colhead{Exposure (s)}
 & \colhead{Airmass}
 & \colhead{S/N per pix\tablenotemark{b}}
}
\startdata
56681.5 & -8.9 & GAO 1.5m & GAOES & $\sim 37,000$ & $4,800 -6,700$ & $1,800 \times 6$ & 1.5 & $\sim 60$\\
56684.6 & -5.8 & GAO 1.5m & GAOES & $\sim 37,000 $ & $4,800-6,700$ & $1,800 \times 8$ & 1.2 & $\sim 100$\\
56696.6 & +6.2 & OAO 1.88m & HIDES & $\sim 50,000$ & $5,000-7,500$ & $1,800 \times 2$ & 1.3 & $\sim 45$\\
56703.5 & +13.1 & Subaru 8.2m & HDS & $\sim 50,000$ & $3,700-6,400$ & $1,800 \times 1$ & 1.6 & $\sim 170$\\
56713.6 & +23.2 & OAO 1.88m & HIDES & $\sim 50,000$ & $5,000-7,500$ & $1,500 \times 4$ & 1.2 & $\sim 30$\\
56735.5 & +45.1 & Subaru 8.2m & HDS & $\sim 50,000$ & $3,700-6,400$ & $900 \times 6$ & 2.5 & $\sim 150$\\ 
& & & &  & $6,400 - 8,500$ & $2,400 \times 1$ &  & \\ 
56748.5 & +58.1 & GAO 1.5m & GAOES & $\sim 37,000$ & $4,800-6,700$ & $1,800 \times 5$ &1.2 & $\sim 25$\\
56755.5 & +65.1 & GAO 1.5m & GAOES & $\sim 37,000$ & $4,800-6,700$ & $1,800 \times 8$ & 1.2 & $\sim 35$\\
56945.5 & +255.1 & Subaru 8.2m & HDS & $\sim 40,000$ & $3,700-6,400$ & $1,800 \times 4$ & 2.8 & $\sim 30$
\enddata
\tablenotetext{a}{GAO (Gunma Astronomical Observatory); OAO (Okayama Astrophysical Observatory); Subaru (Subaru Telescope).}
\tablenotetext{b}{At the continuum near \ion{Na}{1}~D. }
\end{deluxetable*}

\begin{figure*}
\begin{center}
        \begin{minipage}[]{0.8\textwidth}
                \epsscale{1.0}
                \plotone{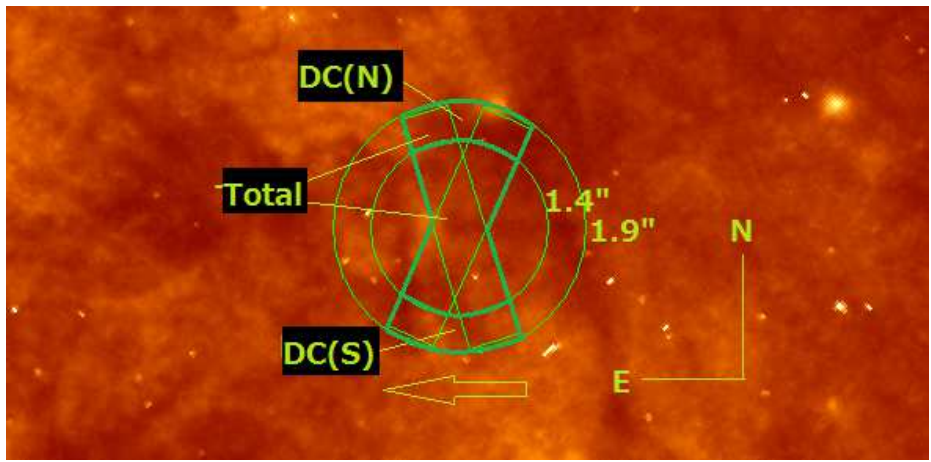}
        \end{minipage}
\end{center}
\caption{Pre-explosion image of the position of SN~2014J obtained
  with {\it HST}/ACS/WFC in the F555W filter (Obs Id: 10776, PI:
  Mountain).  The SN position \citep{crotts2014} is at the center of
  the green circles, which have radii of 1.4 and 1.9\arcsec.  The slit
  orientation for the $+255.1$~day spectrum is indicated by the green
  rectangles -- the slit PA was rotated from 197$^{\circ}$ to
  156$^{\circ}$ during the observation. For our fiducial extraction
  (`total' spectrum), the region along the slit within an aperture of
  1.9\arcsec (within the outer circle) is integrated. The
  representative diffuse component spectra are extracted within the
  region between 1.4\arcsec\ and 1.9\arcsec\ (the intersection of the
  annulus and the rectangles) on both sides of the SN position (N and
  S), which are further used in extracting a `pure' SN spectrum in the
  DC model 2.\label{fig1}}
\end{figure*}

In this paper, we present a series of high-dispersion spectra of
SN~2014J, covering multiple epochs from $-8.9$ to $+255.1$~days
relative to $B$-band maximum brightness, collected by three
telescopes. Especially unique is our latest spectrum at $+255.1$~days;
it is the first high-resolution spectrum taken for any SN~Ia at a
late-time, nebular phase. This paper focuses on unique information we
obtained based on this spectrum, which is qualitatively different from previous
analyses performed for high-resolution spectra of SNe. In \S 2, we
summarize observations and our standard data reduction procedures.
Results are described in \S 3. The paper is closed in \S 4 with
conclusions and discussion. Additional analyses are presented in the
appendices.

\section{Observations and Data Reduction}

\begin{figure*}
\begin{center}
        \begin{minipage}[]{0.9\textwidth}
                \epsscale{0.8}
                \plotone{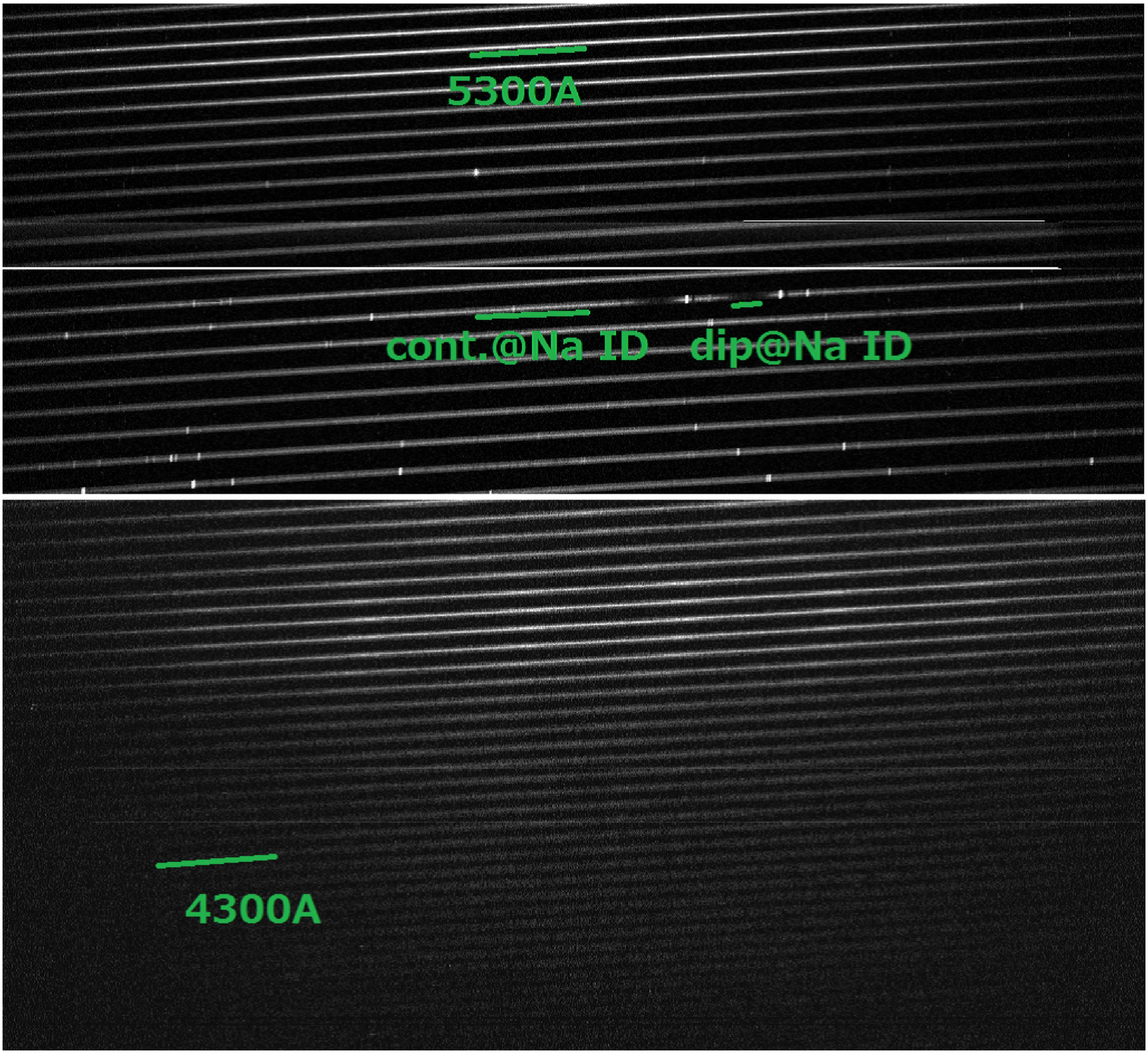}
        \end{minipage}
\end{center}
\caption{Two-dimensional spectrum of SN~2014J at $+255.1$~days
  relative to $B$-band maximum brightness. Spatial profiles (Figure~4)
  are extracted in the region marked in the figure.
  \label{fig2}}
\end{figure*}

\begin{figure*}
\begin{center}
        \begin{minipage}[]{0.9\textwidth}
                \epsscale{0.8}
                \plotone{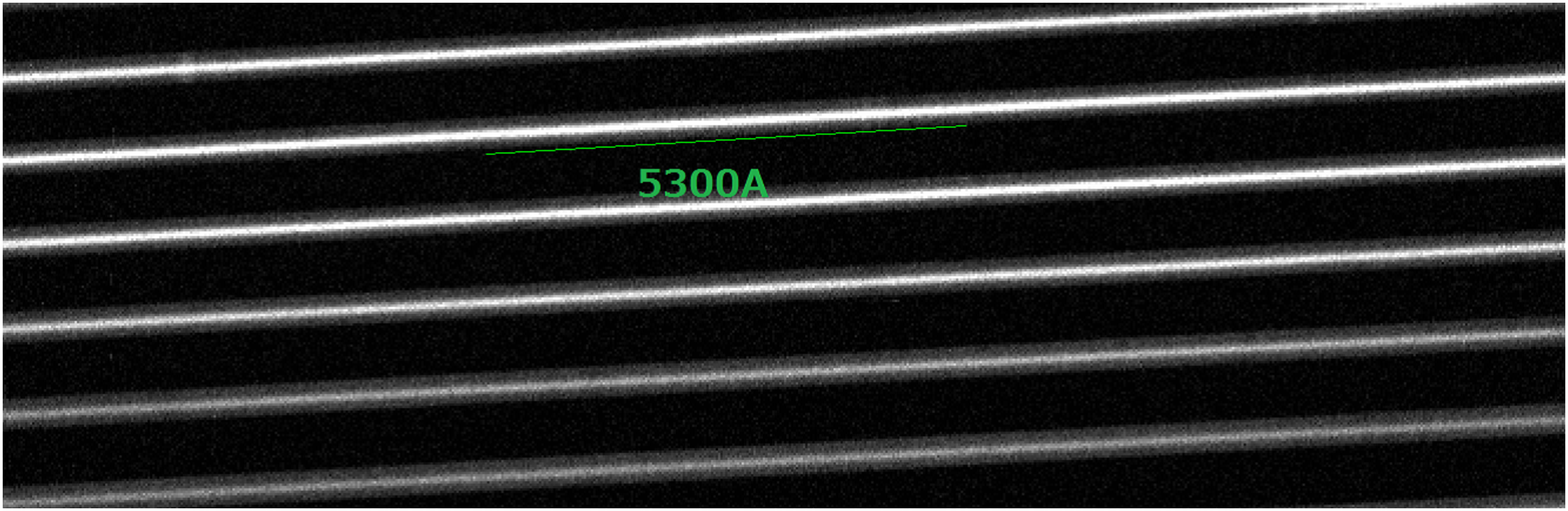}
        \end{minipage}
\end{center}
\caption{Expanded view of the two-dimensional spectrum (Figure~2) near
  $5300$\AA, a region where the SN flux is large relative to the
  background.  \label{fig3}}
\end{figure*}

A log of our high spectral resolution observations is given in
Table~1. We used three telescopes to obtain the spectra, starting at
$-8.9$~days relative to $B$-band maximum brightness \citep[MJD
56690.4:][]{kawabata2014} and ending at $+255.1$~days\footnote{We note
  that the time of maximum adopted in this paper
  \citep[from][]{kawabata2014} is slightly later (by $\sim$0.6~days) 
  than that presented by \citet{marion2015a} (MJD $56689.8 \pm 0.1$)
  and some other works. The small difference ($<$1~day) here would not
  affect any of our analyses and conclusions.}: the 1.5-m telescope at
the Gunma Astronomical Observatory (GAO) using the GAO Echelle
Spectrograph (GAOES), the 1.88-m telescope at the Okayama
Astrophysical Observatory (OAO) of National Astronomical Observatory
of Japan (NAOJ) with the HIgh Dispersion Echelle Spectrograph (HIDES)
in a fiber-feed mode \citep{kambe2013}, and the 8.2-m Subaru telescope
of NAOJ equipped with the High Dispersion Spectrograph (HDS)
\citep{noguchi2002}. The observational parameters (e.g., spectral
resolution and wavelength coverage) are shown in Table~1, where the
spectral resolution and the signal-to-noise ratio (S/N) at the continuum around
\ion{Na}{1}~D are measured from the reduced spectra.

We followed standard IRAF\footnote{IRAF is distributed by the National
  Optical Astronomy Observatory, which is operated by the Association
  of Universities for Research in Astronomy, Inc., under cooperative
  agreement with the National Science Foundation.} procedures to
reduce the data. Scattered light was subtracted but no further
`astronomical' background subtraction was performed in our standard
reduction procedure (i.e., generally a typical procedure for
high-resolution data reduction). The wavelength calibration was
performed using observations of Th-Ar lamps, resulting in an accuracy
of $\sim$0.0063~\AA\ for the case of the spectrum at $+255.1$ days.  A
heliocentric velocity correction was applied to each spectrum, and the
wavelength was further redshifted to the M82 rest frame ($z =
0.000677$).

\begin{figure}
        \begin{minipage}[]{0.45\textwidth}
                \epsscale{1.2}
                \plotone{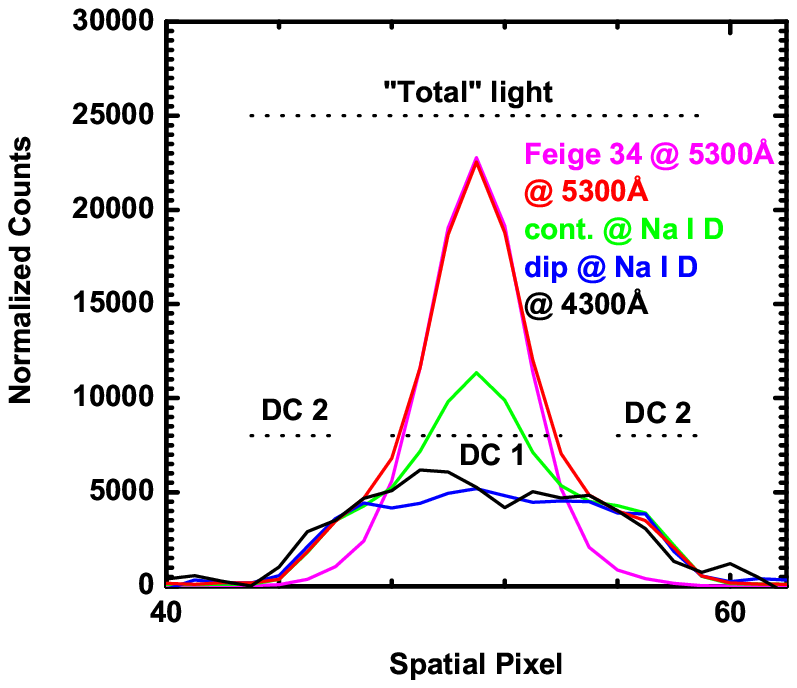}
        \end{minipage}
\caption{Spatial profiles of the $+255.1$~day SN~2014J spectrum at
  different wavelengths: at $\sim$5300~\AA\ where the SN light
  dominates (red), the region corresponding to the continuum near
  \ion{Na}{1}~D where the SN and underlying diffuse light have
  comparable contributions (green), a region corresponding to strong
  \ion{Na}{1}~D absorption (blue) and the continuum at $\sim$4300~\AA\
  where the diffuse component dominates over the SN (black).  One
  spatial pixel in this figure has an angular size of
  $\sim$0.28\arcsec\ (after the binning before the CCD readout). The
  fluxes are arbitrarily normalized to provide comparable fluxes in
  the `diffuse' component at different wavelengths. For comparison,
  the spatial profile of a spectrum of the standard star Feige~34 is
  shown for the $\lambda \sim 5300$~\AA\ (magenta). \label{fig4}}
\end{figure}

To produce continuum-normalized flux spectra, first the continuum was
fit for each order separately, and the resulting blaze function was
then applied to the original spectra when convolving the spectra from
different orders. To obtain the continuum-normalized sky spectra, the
same procedures were performed for telluric standard stars. The
continuum-normalized SN spectra were then divided by the
continuum-normalized sky spectra to remove telluric absorption. We did
not remove the Milky Way (MW) absorption features in the sky spectra,
and this could induce the appearance of variable MW features; however,
we do not believe any such variability is real.

Additionally, we created a flux-calibrated version of the $+255.1$~day
spectrum. A flux standard star, Feige~34 \citep{oke1990}, was observed
during the same night with the same settings, and the sensitivity
function was derived for each order, which we applied to the SN
spectrum.  The sky subtraction was not performed for this spectrum,
but it does not affect our results derived from the flux-calibrated
spectrum.

In this paper, we will mostly focus on implications obtained from the
$+255.1$~day spectrum; here we provide a more detailed description of
its observation and data reduction. Figure~1 shows our observational
setup (slit size and angle) for this observation, overlapped with a
pre-explosion image taken by {\it HST}/ACS/WCS in the F555W filter
\citep[Obs ID: 10776, PI: Mountain, see also][]{crotts2014}. For 
this observation, we did not use an image rotator, but did employ an
atmospheric dispersion corrector. The slit position angle (PA) was rotated from 197 deg
to 156 deg during the sequence of exposures that total $7,200$ sec.
Namely, on average the slit direction was north (N) - south (S), but
we collected the light within the region extending to $\pm$20 degree
in PA.

Four exposures of $1800$~sec each were obtained for the $+255.1$~day
spectrum. The airmass was high, varying from 2.8 to 2.1. In
Appendix~A, we checked the possibility that the high airmass might
introduce any systematic effects, and we conclude that such a
possibility is highly unlikely. The 2D spectrum from $+255.1$~days
(with the four exposures combined) is shown in Figure~2, and an
expanded view around the wavelength region where the SN flux is high
($\sim$5300~\AA) is shown in Figure~3. The 2D spectrum shows that the
SN is clearly detected.  Figure~4 shows the spatial distribution of
the incoming light for different wavelength regions (see Figure~2). We
see two components, one with a FWHM (full width at half-maximum) of
$\sim$0.8\arcsec\ whose flux strongly depends on the wavelength, and
another component extending through the entire aperture whose spatial
profile is not sensitive to the wavelength. For comparison, the
  spectrum of the flux standard star (Feige~34) taken on the same
  night shows only a single component without an extended component
  (Figure~4). In the SN spectrum, there is a clear trend that the
  first component is seen only in the wavelength range where the SN is
  expected to be bright -- for example, at $\sim$4,300~\AA\ the flux
  is low (Figure~2) and indeed only the second, extended component is
  identified (Figure~4). We identify the former as the SN
contribution and the latter as diffuse light from the region near
SN~2014J in M82 (see \S 3 for further details). In our default
spectral extraction, the spectrum was extracted within the whole
aperture, which is $0.8 \times 3.8$ arcsec$^{2}$ (shown as `total' in
Figures~1 and 4). This turns out to be important, and we will revisit
this issue in \S 3.

\section{Results}

\subsection{Variable \ion{Na}{1}~D Systems at Late Times?} 

\begin{figure*}
\begin{center}
        \begin{minipage}[]{0.45\textwidth}
                \epsscale{1.0}
                \plotone{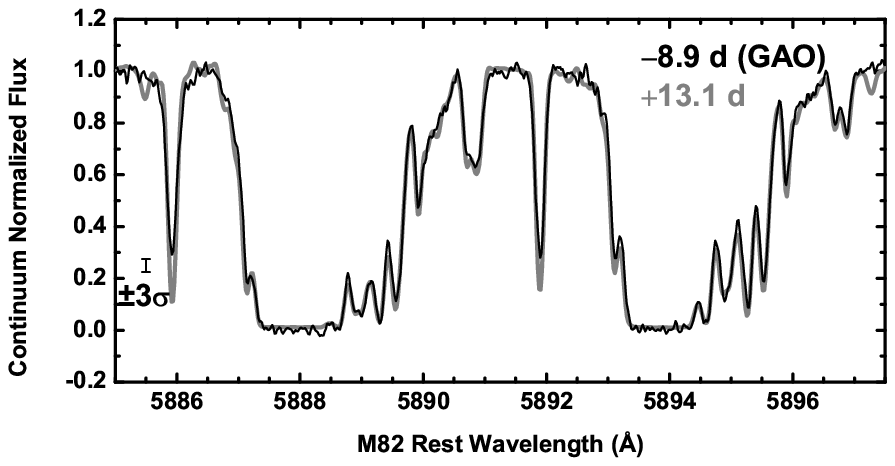}
        \end{minipage}
         \begin{minipage}[]{0.45\textwidth}
                \epsscale{1.0}
                \plotone{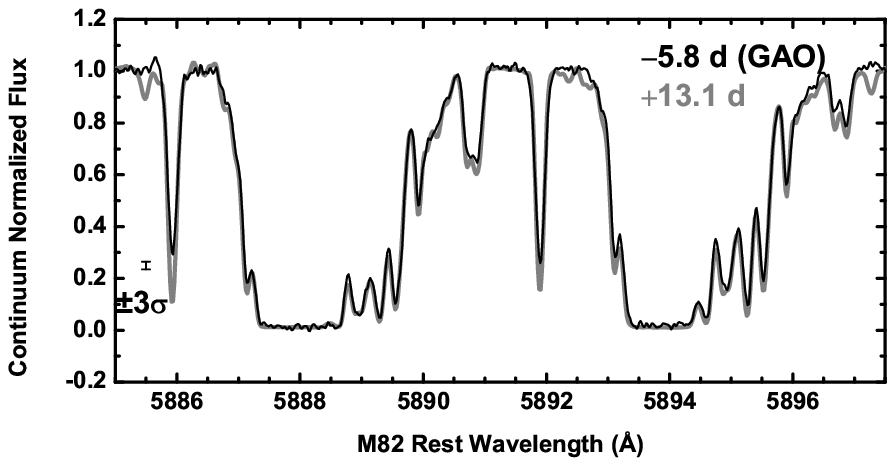}
        \end{minipage}\\
         \begin{minipage}[]{0.45\textwidth}
                \epsscale{1.0}
                \plotone{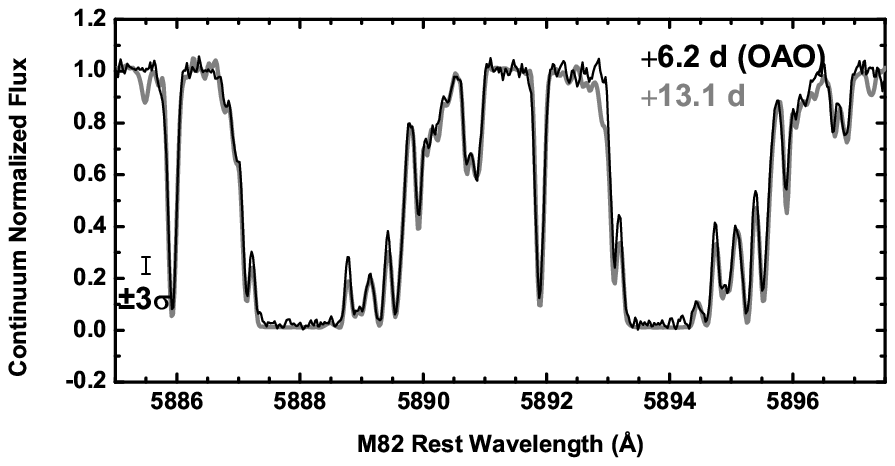}
        \end{minipage}
         \begin{minipage}[]{0.45\textwidth}
                \epsscale{1.0}
                \plotone{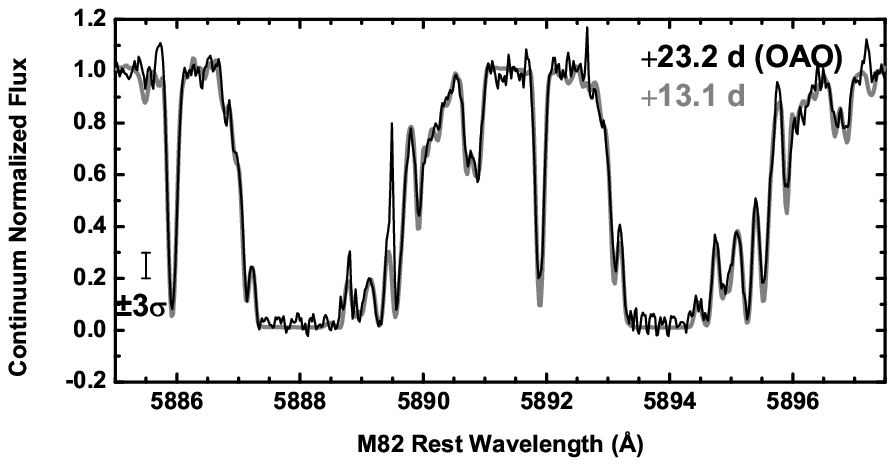}
        \end{minipage}
         \begin{minipage}[]{0.45\textwidth}
                \epsscale{1.0}
                \plotone{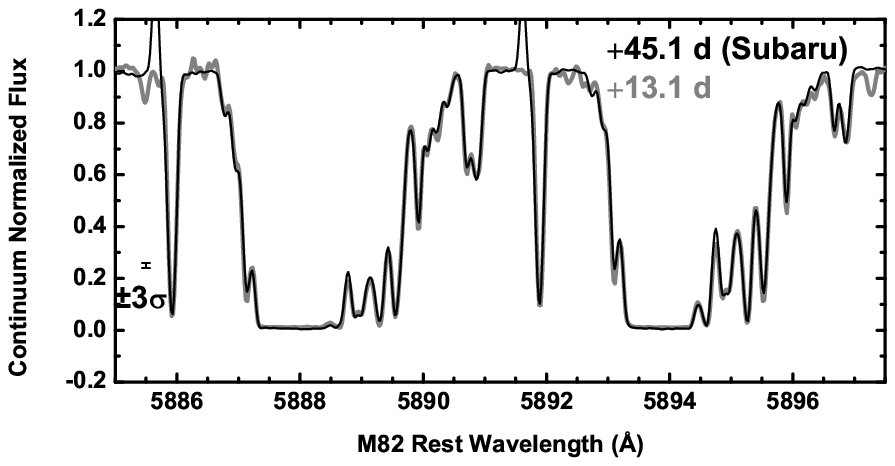}
        \end{minipage}
         \begin{minipage}[]{0.45\textwidth}
                \epsscale{1.0}
                \plotone{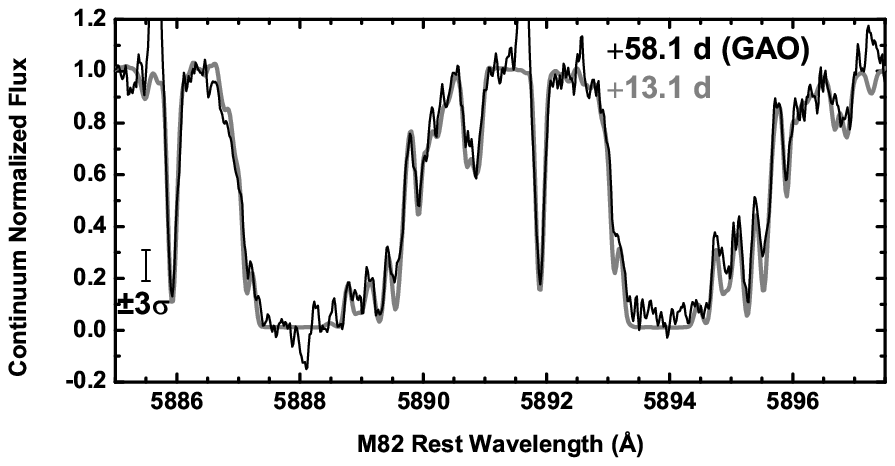}
        \end{minipage}
         \begin{minipage}[]{0.45\textwidth}
                \epsscale{1.0}
                \plotone{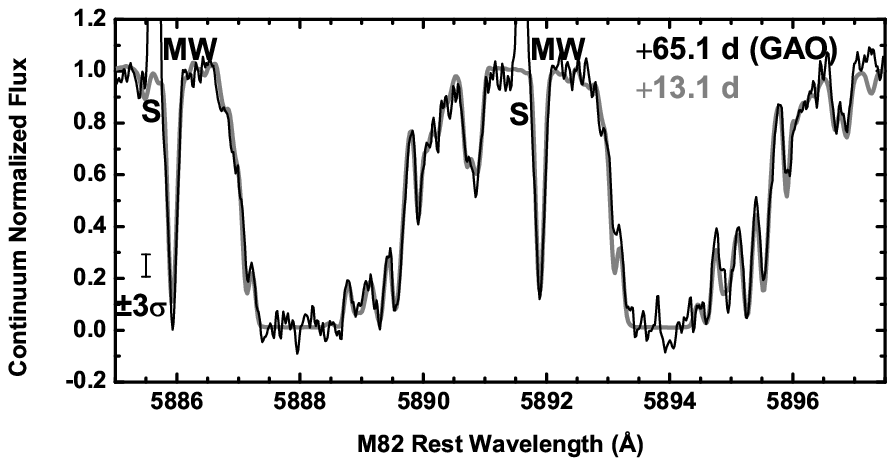}
        \end{minipage}
         \begin{minipage}[]{0.45\textwidth}
                \epsscale{1.0}
                \plotone{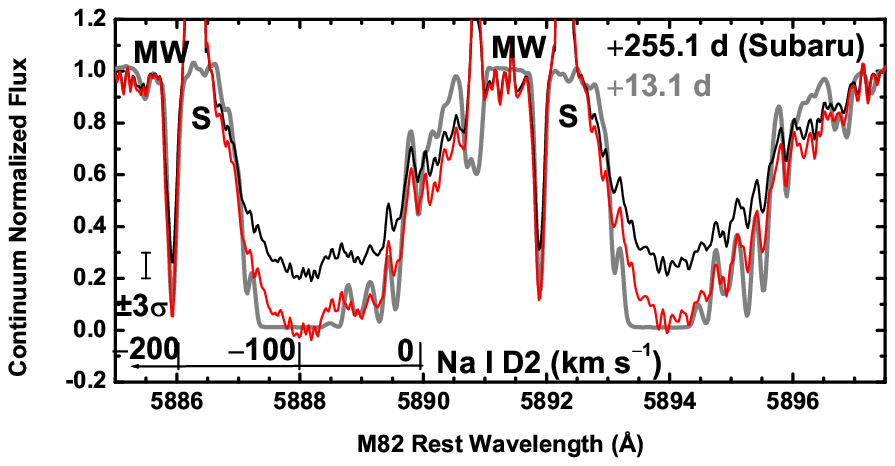}
        \end{minipage}
\end{center}
\vspace{-1cm}
\caption{Series of high-resolution spectra of SN~2014J (black). The
  spectra are sky-subtracted using telluric standard stars (see \S 2
  for details). In each panel, the spectrum at $+13.1$~days taken by
  the Subaru telescope is shown for comparison (gray), and is
  convolved with a Gaussian kernel to provide a comparable spectral
  resolution for each spectrum. For the $+255.1$~day spectrum, the
  original spectrum (black) shows an offset in the base of the
  (previously) saturated components clearly exceeding the noise level.
  A spectrum after subtracting a constant flux is shown in red. A
  velocity scale for \ion{Na}{1}~D2 in the M82 rest frame is indicated
  in the last panel.  The night-sky emissions of \ion{Na}{1}~D1 and D2
  are marked by `S' and the Milky Way components are marked by `MW' in
  the lower two panels.
  \label{fig5}}
\end{figure*}

Figure~5 shows our high-resolution spectral series of SN~2014J,
focusing on the wavelength range covering \ion{Na}{1}~D1 (5895.92~\AA)
and D2 (5889.95~\AA). The first five spectra ($-8.9$ to $+23.2$~days)
confirm the result from the previous works that the \ion{Na}{1}~D
absorption systems does not show significant variations up to
+30.4~days after maximum brightness \citep{foley2014, ritchey2015,
  graham2014}. We find that the lack of detected variability is
further extended until $+65.1$~days, where the high-S/N $+45.1$~day
spectrum (${\rm S/N} \approx 150$ per pixel) strongly constrains
variability while the lower-S/N $+58.1$ and $+65.1$~day spectra (${\rm
  S/N} \approx 30$ per pixel) are less constraining.

The $+255.1$~day spectrum, the first high-resolution spectrum of any
SN during the nebular phase, has \ion{Na}{1}~D profiles with a
different appearance to those from earlier phases. The (previously)
saturated component around $-100$~km~s$^{-1}$ (in the M82 rest frame)
has non-zero flux in this spectrum. This flux level is beyond
$3\sigma$ and also this feature is not affected by the treatment of
the `scattered light' subtraction using the fluxes in the regions
between different orders in the 2D spectrum (Figure~2). These
arguments suggest that the non-zero flux is real and has an
astrophysical origin. Furthermore, even if we subtract a constant flux
across the wavelength on the assumption that the previously saturated systems
should remain so in our +255.1 day spectrum, the resulting spectrum
shows noticeable differences to the earlier-phase spectra, especially
visible in the bluer and redder parts of the saturated component
($5886.5-5887.5$\AA\ and $5890.0-5890.5$\AA\ for \ion{Na}{1}~D2).  We
therefore conclude that this difference is real.

The question we seek to answer in this paper is whether this behavior in
the late-time spectrum indicates time variability in the \ion{Na}{1}~D
systems toward SN~2014J, or if not, how this can be understood. This
is then connected to a general question about how and where the
absorption lines are created, especially if one can identify any CSM
components in the narrow absorption-line systems. In the following, we
will first show that the feature as mentioned above is indeed caused
by contamination with diffuse light coming from regions not
exclusively aligned with the line-of-sight toward the SN. Using this
information, we will show that all the absorption components seen in
the high-resolution spectra of SN~2014J originate in the foreground
region having a physical scale of $\gsim $40~pc, and thus the
existence of CSM components in the observed \ion{Na}{1}~D absorption
systems is rejected for this particular SN.

\subsection{The Size and Position of the Absorbing Systems}

\begin{figure*}
\begin{center}
        \begin{minipage}[]{0.7\textwidth}
                \epsscale{1.0}
                \plotone{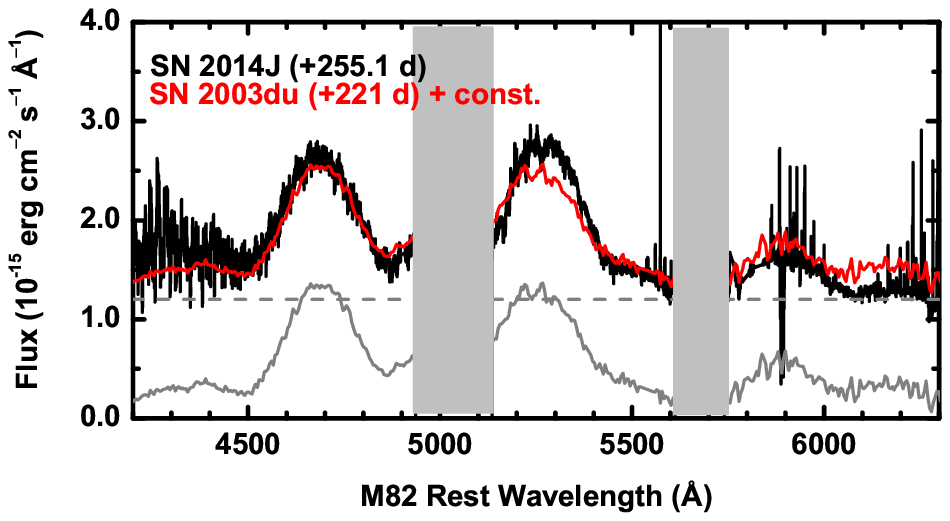}
        \end{minipage}
        \begin{minipage}[]{0.7\textwidth}
                \epsscale{1.0}
                \plotone{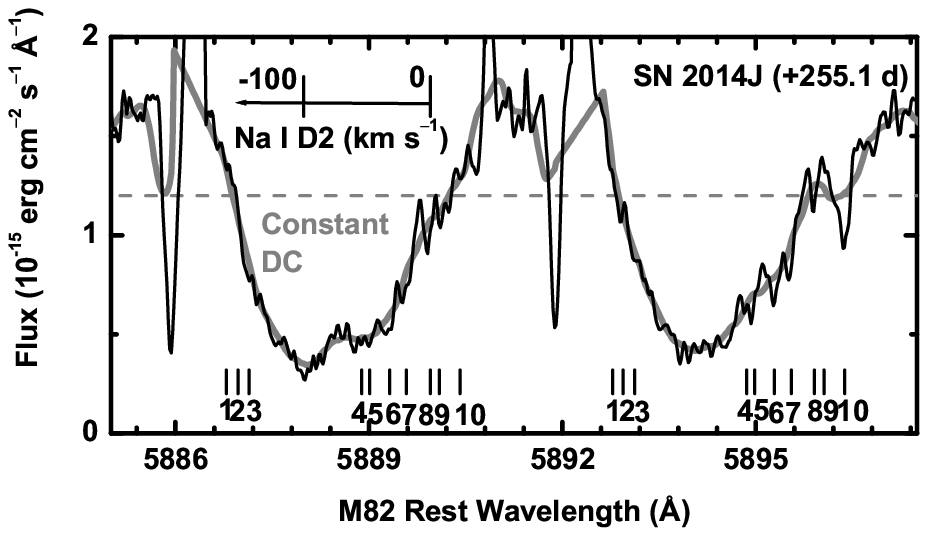}
        \end{minipage}
\end{center}
\vspace{-1cm}
\caption{(Upper panel) Flux-calibrated spectrum of SN~2014J at 
  $+255.1$~days relative to $B$-band maximum brightness, smoothed with
  a 25~pixel boxcar filter (black). The sky (telluric) lines are not
  subtracted. CCD gaps are masked by the gray regions. For comparison,
  a spectrum of the normal SN~Ia~2003du at $\sim$221~days relative to
  $B$-band maximum brightness is also shown \citep{stanishev2007},
  for which the wavelength and the flux are brought to the M82 rest
  frame (gray, solid). The spectrum of SN~2013du is reddened to match
  the total extinction toward SN~2014J, $E(B-V) = 1.37$~mag and
  $R_{V} = 1.4$. Adding a constant $f_{\lambda}$ continuum (gray,
  dashed) to the SN~2003du spectrum, the sum (shown in red) is similar
  to the `total' SN~2014J spectrum. (Bottom panel) The same SN~2014J
  spectrum focusing on the wavelength region around \ion{Na}{1}~D
  (gray).  Also shown is the unsmoothed spectrum (black). The dashed
  line labeled as `constant DC' is the assumed continuum level in the
  diffuse component, the same as that shown in the upper panel. 
  A velocity scale for \ion{Na}{1}~D2 in the M82 rest
  frame is indicated in the lower panel.
\label{fig6}}
\end{figure*}

Figure~6 shows the flux-calibrated $+255.1$-day spectrum of SN~2014J,
following our fiducial reduction procedures. For comparison, the
positions of the 10 unsaturated components covering the velocity range 
from $-160$ to $+23$~km~s$^{-1}$ in the M82 rest frame, as identified
by \citet{graham2014}, are shown. One immediately notices that the
spectrum is indeed contaminated by a component (or components) other
than the SN itself: At this late-phase, the SN emission is created
within the optically thin ejecta, characterized by forbidden lines
(from Fe-peak elements for SNe~Ia) and virtually zero continuum flux.
In Figure~6, this is seen in the spectrum of normal SN~Ia~2003du
\citep{stanishev2007} but reddened using $E(B-V) = 1.37$~mag and
$R_{V} = 1.4$ as derived for SN~2014J \citep{goobar2014,
  kawabata2014}. Adopting slightly different values for the extinction
\citep{foley2014} does not change our conclusions. After correcting
for the difference in the distances to and extinction within the host
galaxies, it is clear that the spectra are quite similar in both the
general features and the flux scale, except that there is an offset in
the level of continuum. If we add a (roughly) constant flux across the
observed wavelength range to the spectrum of SN~2003du, the overall
feature of the late-time spectrum of SN~2014J is well reproduced. This
shows that nothing peculiar is taking place in our late-time spectrum,
but rather that it can simply be interpreted as a sum of the SN and the
light from background (or foreground) regions (hereafter the SN-site
diffuse component). Note that the flat continuum implies an observed
color of $B-V \approx 0.6$~mag, which is roughly consistent with the
color of diffuse light in M82 \citep{mayya2009}.

The $+255.1$~day spectrum was extracted such that the aperture was
chosen to be roughly the slit length, i.e., corresponding to
$\sim$2\arcsec\ of radius (Figures~1 and 4). This translates into a
projected physical scale of $\sim$40~pc at the distance of M82 (or
$\sim$80~pc in diameter). For comparison, the SN ejecta must have
expanded to $\sim$0.02~pc for the ejecta velocity of $\sim$0.1$c$, and
the emitting region should be even smaller than this since the light
is emitted from the innermost region in the nebular phase. On the
other hand, there is a substantial contamination from the diffuse
light in our integrated $+ 255.1$~day spectrum, becoming $\sim$70\% of
the total light at the wavelength of \ion{Na}{1}~D (Figure~6). Only a
negligible fraction of this diffuse light should subtend the same
angle subtended by the SN ejecta, while overall the diffuse light
shows the similar \ion{Na}{1}~D systems to those seen toward SN~2014J
in the earlier phases (when contamination by the diffuse light was
negligible).

While typically the SN-site diffuse flux can be (and is) neglected for
high-resolution observations of SNe (taken around maximum brightness),
this is not the case for the last epoch in our observations: the SN
flux has decreased substantially at $+255.5$~days, and the SN is in
an active starburst galaxy, M82. Indeed, substantial contamination
from M82 is inferred from the spatial distribution of the incoming
light (Figure~4). A component with a FWHM of $\sim$0.8\arcsec\ is
clearly identified as coming from a point source, and this component
is superimposed by another component with a more extended spatial
distribution covering at least the length of the slit.  The flux
ratio between the two components depends on the wavelength, and the
contribution from the point source is large where strong emission
lines exist in SN~Ia spectra at a similar epoch (Figures~4 \& 6).
Therefore, the point source component is identified as the SN while
the diffuse component is likely coming from the diffuse light of M82.
This interpretation is further tested below.

\begin{figure*}
\begin{center}
        \begin{minipage}[]{0.8\textwidth}
                \epsscale{1.0}
                \plotone{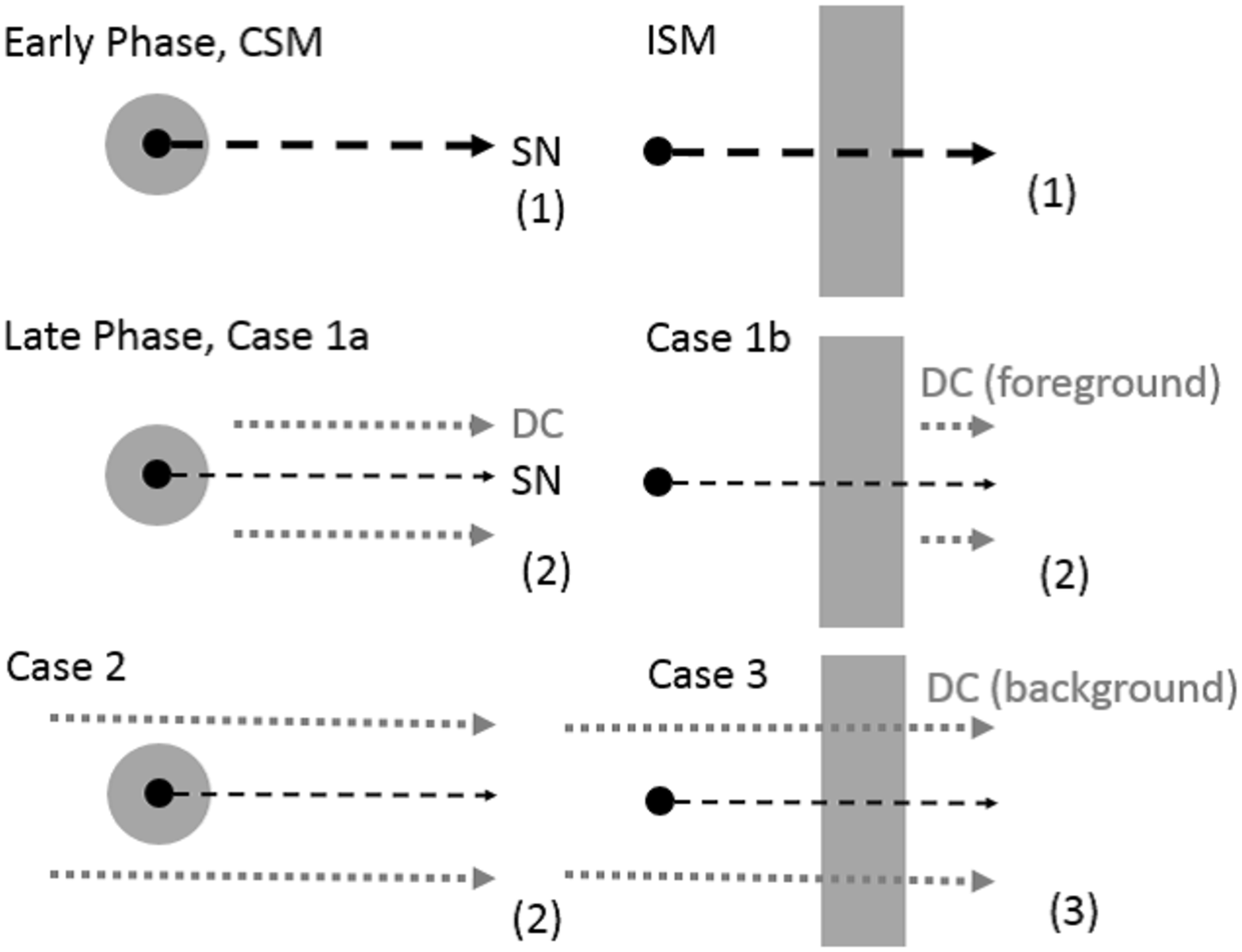}
        \end{minipage}
        \begin{minipage}[]{0.3\textwidth}
                \epsscale{1.0}
                \plotone{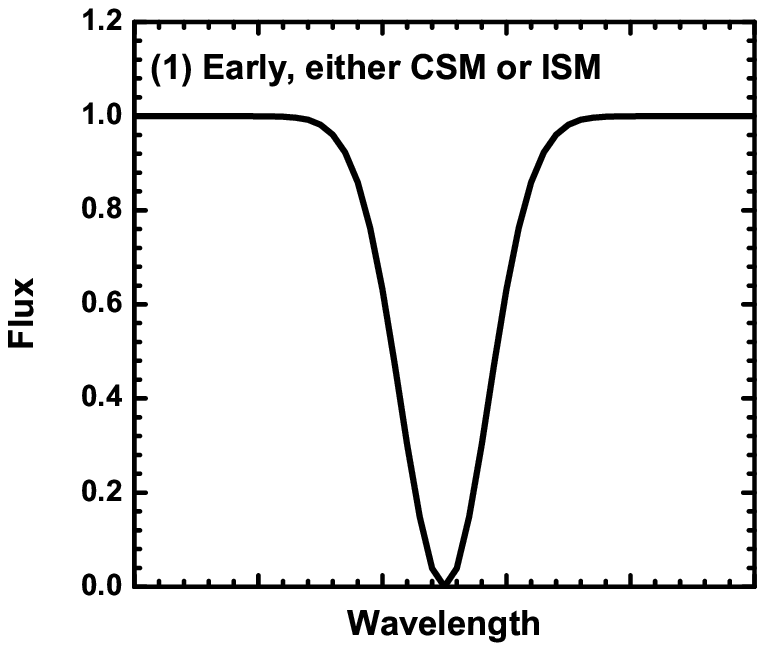}
        \end{minipage}
        \begin{minipage}[]{0.3\textwidth}
                \epsscale{1.0}
                \plotone{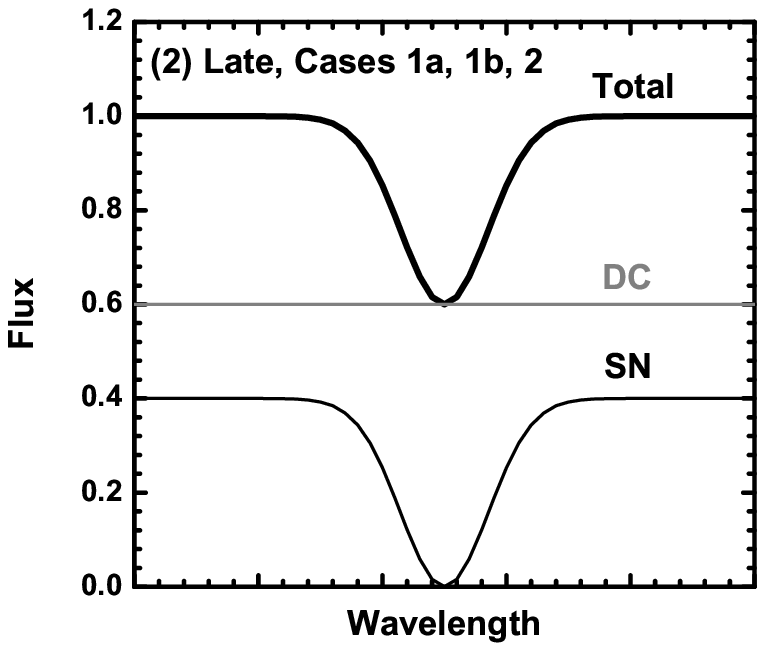}
        \end{minipage}
        \begin{minipage}[]{0.3\textwidth}
                \epsscale{1.0}
                \plotone{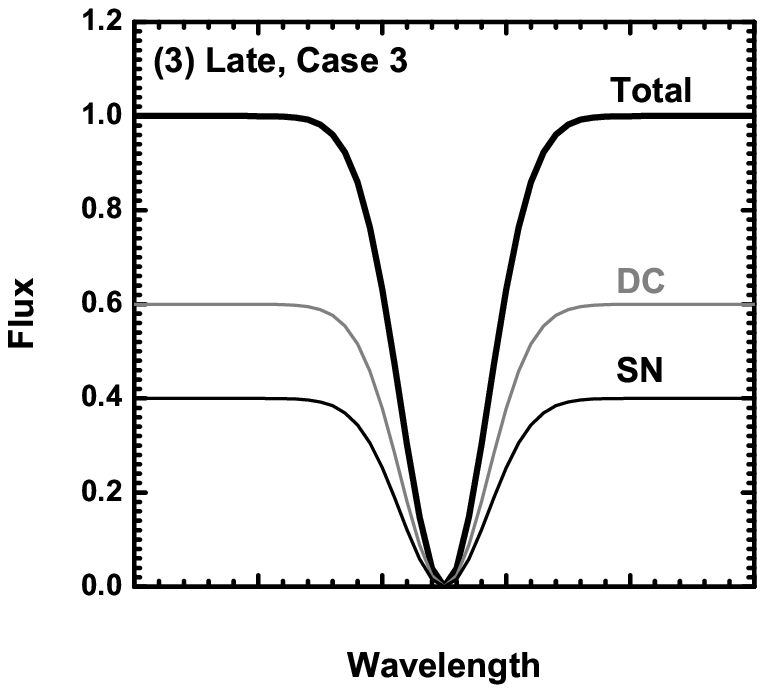}
        \end{minipage}
\end{center}
\vspace{-1cm}
\caption{Sketch of the \ion{Na}{1}~D absorbing systems toward a
  point source (SN) with a contribution from an unrelated diffuse
  component (DC). An example is shown for a single line that is
  saturated through the absorbing system under consideration, either
  CSM or ISM. In the early phase, the SN is bright and the
  contribution from the DC is negligible, and the spectrum is formed
  by light that goes through the absorbing system as shown in (1). If
  a large fraction of the DC does not pass through the absorbing
  system, the absorption-free DC creates the non-zero floor at the
  bottom of the absorption line at the flux level of the DC even if
  the line is saturated within the absorbing system, as shown in (2).
  Finally, if the DC shares the same absorbing system as the SN
  light, the (flux-normalized) spectra of these two components are the
  same, and thus the total spectrum, being a combination of the two, is
  identical to that observed in the early phase, as shown in (3).
  \label{fig7}}
\end{figure*}

We have checked the surface brightness of the region around the SN in
a pre-SN image taken with {\it HST}/ACS/WCS in the F555W filter (i.e.,
covering the region containing \ion{Na}{1}~D), as shown in Figure~1
\citep[Obs ID: 10776, PI: Mountain, see also][]{crotts2014}. We
estimate that the surface brightness is $\sim$18.3~mag~arcsec$^{-2}$.
Indeed, this is consistent with the typical value of the surface
brightness, $\sim$18.5~mag~arcsec$^{-2}$, in the $V$-band at the
position $\sim$60\arcsec\ away from the center of M82 along the major
axis \citep[e.g., ][]{mayya2009}. If we integrate this diffuse light
within the aperture of $0.8 \times 3.8$~arcsec$^{2}$ (corresponding to
the slit size), a nominal value in our spectral extraction, the
contaminating light would result in $V \approx 17.1$~mag, comparable
to the expected magnitude of SN~2014J at this phase extrapolated from
the early-time light curve (i.e., $V \approx 17$~mag). By performing
spectrophotometry on our fluxed spectrum, we obtain $B \approx
16.4$~mag and $V \approx 16.5$~mag, values consistent with the idea
that the spectrum is a sum of the SN and the diffuse light with
comparable contributions. These arguments are also consistent with the
spatial distribution of the incoming light (Figure~4). We note that
this photometry is not very accurate, since this procedure assumes
point-like sources for both the flux standard star and the target,
which is not true for the substantial contribution from the diffuse
light. In any case, none of our arguments relies on the absolute flux
scale, and thus it does not affect any of our conclusions.

Importantly, while there is non-zero flux at the bottom of the
`saturated' components in \ion{Na}{1}~D (as shown in Figure~5 as
well), the bottom of the \ion{Na}{1}~D features are {\it below} the
SN-site diffuse light.  This is strong evidence that
the SN-site diffuse light is also absorbed by the {\it same} absorbing
features that cause the SN~2014J \ion{Na}{1}~D profiles (see below). We note that
a constant flux of the diffuse continuum light is merely a first-order
approximation, and indeed it is better fit by a bluer continuum with
the $B$-band flux larger than the $V$-band by $\sim$20--30\%. We have
checked several continuum models for the diffuse light (constant,
linear, parabolic) and confirmed that the uncertainty here would not
change our arguments --- in all cases the depths of absorptions exceed
the continuum level.

\begin{figure*}
\begin{center}
        \begin{minipage}[]{0.7\textwidth}
                \epsscale{1.0}
                \plotone{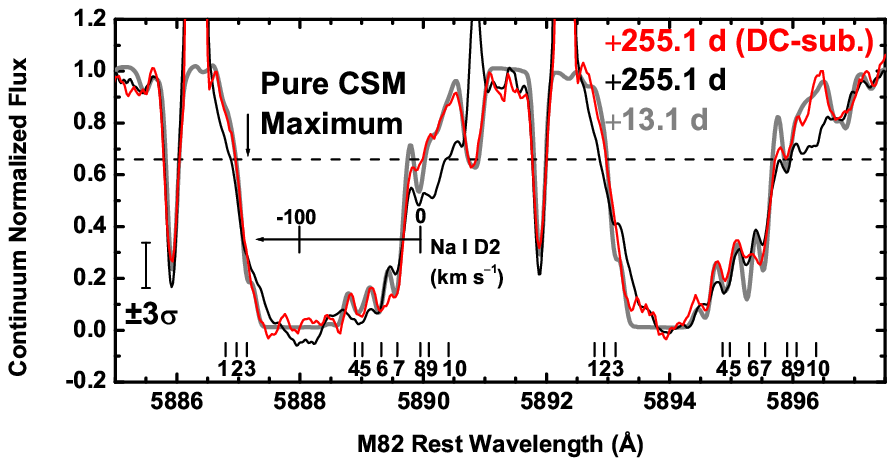}
        \end{minipage}
        \begin{minipage}[]{0.7\textwidth}
                \epsscale{1.0}
                \plotone{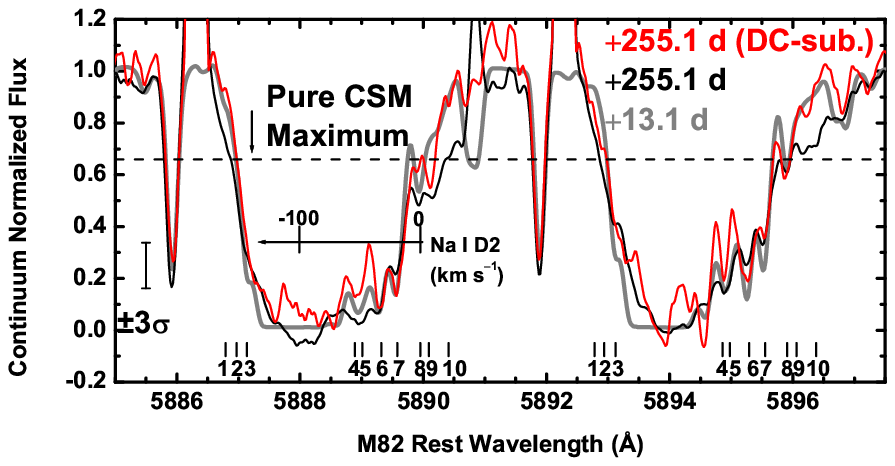}
        \end{minipage}
\end{center}
\vspace{-1cm}
\caption{`Contamination-subtracted,' i.e., the `pure SN,'
  high-resolution $+255.1$~day spectrum of SN~2014J (red). We also
  display the total-light spectrum where we have simply subtracted a
  constant flux spectrum (black) and the $+13.1$~day spectrum (gray).
  For the aesthetic reasons, all spectra are smoothed with a 7 pixel
  boxcar filter. Two models for the underlying diffuse component are
  tested: In Model 1 (upper panel), the off-SN region (at
  $\sim$0.8\arcsec) was subtracted simultaneously when extracting the
  1D SN spectrum. In Model 2 (lower panel), the 1D spectrum of the
  off-SN diffuse component was extracted independently from the SN
  spectrum extraction within the annuli of 1.4 -- 1.9\arcsec, and then
  the flux of this diffuse light spectrum is scaled to fit to the
  minimum flux of the unsubtracted SN spectrum at the bottom of the
  (presumably) saturated \ion{Na}{1}~D components. The resulting
  contaminating spectrum is then subtracted from the original
  spectrum. The dashed line marked `Pure CSM Maximum' indicates the
  maximum depth for a component coming exclusively from the CSM (i.e.,
  along a pencil-like beam toward the SN), where the putative CSM
  component is assumed to be saturated without any contribution from
  the larger physical scale (see Figure~7).  This is set by the
  continuum of the DC. \label{fig8}}
\end{figure*}

Depending on whether the origin of the contaminating diffuse light
is in front of the absorbing cloud(s) (i.e., foreground) or behind
it (i.e., background), and on whether the size of the absorbing
cloud(s) is smaller or larger than the size of the SN photosphere, there
are three situations to consider in discussing the depths of the
absorbing components.  This is schematically shown in Figure~7. 
(Case 1: ) The contaminating diffuse light originates in a foreground
region. In this case, the spectrum of the SN-site diffuse light is
free of absorption and would have a roughly flat continuum. Thus, if
the SN line-of-sight absorptions are all saturated, then the {\em
  maximum} depth of the absorbing systems would be at the continuum
level of the contaminating light. (Case 2: ) The contaminating diffuse
light is behind the absorbing system and the absorbing system is
small. In this situation, the contaminating light would not suffer
absorption, and again it is simply a continuum.  Therefore, the same
argument as in Case 1 holds. (Case 3: ) The contaminating diffuse
light is behind the absorbing system, and the absorbing system is
large. In this case, the depths of the absorbing systems are
independent of the contamination of the diffuse light.  Generally, the
contaminating light can be the sum of light originating in the region in
front of the absorbing cloud(s) and light from behind it. However, if the
absorbing cloud is small (a combination of Cases 1 and 2), the depths
of the absorption can never go below the continuum level. Only if the
absorbing cloud is large can the depth of the absorbing system be
below the continuum level, and the depth is basically determined by
the fraction of the light from the region behind the absorbing system
as compared to the foreground light (a combination of Cases 1 and 3).

Therefore, if there are \ion{Na}{1}~D absorbing systems localized
exclusively within a pencil-like beam along the line-of-sight to
SN~2014J, the absorption strengths of such components can never be
below the continuum level of the diffuse component. All the saturated
components are indeed below this level in their fluxes (Figure~6),
immediately rejecting the possibility that any of them is associated
with a local region around the SN. This also applies to most of the
unsaturated components.  Except for the unsaturated components 1 and
10, their depths are below the continuum level. Thus the unsaturated
components 2--9 should not be associated with the local region around
the SN. It is very unlikely that even the remaining components 1 and
10 are from the local SN site. The depths of these components are
comparable to the level of the diffuse light continuum, and therefore
they must be saturated if there is no corresponding absorption system
in the diffuse light. On the other hand, these components were not
saturated in the earlier phases, and indeed they were the weakest
absorbing components among those clearly identified toward SN~2014J.
While this possibility of their being saturated on $+255.1$~days is
not strictly rejected from the argument presented here alone, we will
show in the subsequent sections that this possibility is very
unlikely.

\subsection{Have the Intrinsic SN Absorbing Systems Changed?}
Despite our conclusion that all the visible components of the
\ion{Na}{1}~D systems toward SN~2014J are shared by a diffuse light
from the region extending to $\sim$80~pc in the projected diameter,
this does not readily reject the possibility that some components
might contribute (although not predominantly) via local CSM components
around SN~2014J. The first point to clarify regarding this issue is whether
there is variability in the absorbing spectrum along the `pure'
line-of-sight to SN~2014J. For this purpose, we try to subtract the
contribution from the unrelated diffuse component to extract the pure,
contamination-subtracted SN spectrum. In addition to investigating the
temporal variability in the SN line-of-sight spectrum, this procedure
inversely provides a check on our conclusion that there is a
substantial contamination from the unrelated diffuse light to the SN
spectrum as we will show below.

\begin{figure*}
\begin{center}
        \begin{minipage}[]{0.7\textwidth}
                \epsscale{1.0}
                \plotone{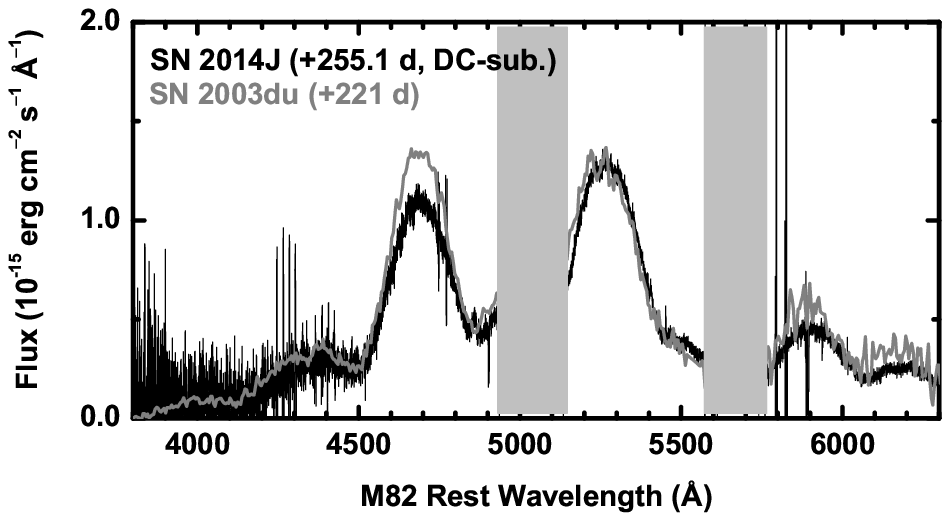}
        \end{minipage}
        \begin{minipage}[]{0.7\textwidth}
                \epsscale{1.0}
                \plotone{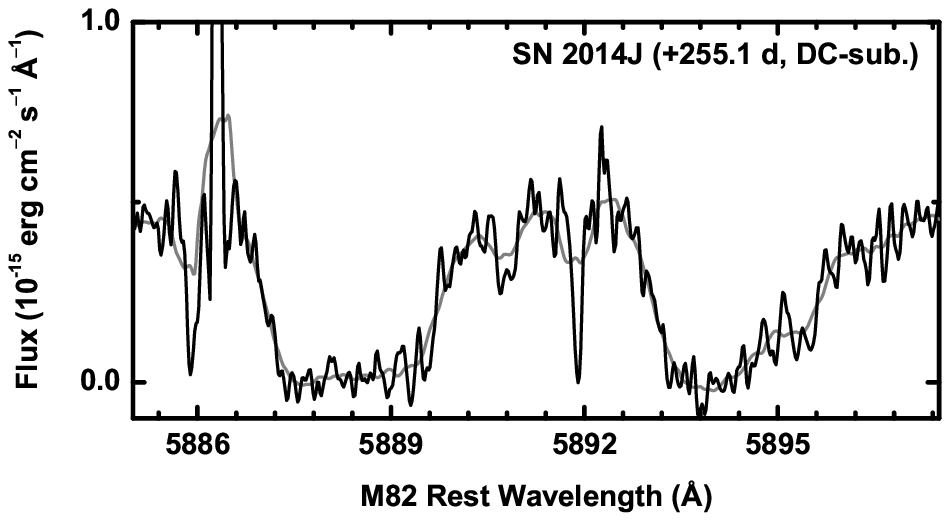}
        \end{minipage}
\end{center}
\caption{Same as Figure~6 (the flux-calibrated spectrum of SN~2014J),
  but after subtracting the diffuse component using the DC model 1. In
  the upper panel, a spectrum of SN~Ia~2003du at $\sim$221~days is
  also shown (gray; distance and extinction ``corrected'' for to match
  those of SN~2014J). In the bottom panel, the smoothed (gray) and
  unsmoothed (black) spectra are shown for the wavelength region
  covering \ion{Na}{1}~D1 and D2.  \label{fig9}}
\end{figure*}

We created a model spectrum of the contaminating diffuse light and
subtracted it from the original target spectrum. For this purpose, we
tested two SN-site contamination models.  In the `DC (diffuse
component) model 1,' the contaminating diffuse light was subtracted
simultaneously when extracting the 1D SN spectrum using the {\tt
  apall} task in IRAF. In doing this, we extracted the SN flux within
an aperture of $\sim$0.8\arcsec\ $\times$ 0.8\arcsec, and the
contaminating light was fit by a linear function connecting the
regions $\sim$0.8\arcsec away from the SN along the slit. In the `DC
model 2', we extracted 1D spectra for regions $\gsim$1.4\arcsec away
from the SN position (in both directions) along the slit (see Figures
1 \& 4).  The spectra of the contaminating diffuse light obtained for
both sides of the target (North and South) were then averaged. The
flux of this contaminating spectrum was scaled to fit to the minimum
flux of the unsubtracted SN spectrum, extracted within
$\sim$1.4\arcsec\ from the SN position, at the bottom of the saturated
\ion{Na}{1}~D components.  Then the contaminating spectrum was
subtracted from the unsubtracted SN spectrum, resulting in a `pure' SN
spectrum.  Note that the contaminating spectrum, in both models, will
contain some SN light (given that the DC extraction regions are relatively
close to the SN position as compared to the seeing size of
$\sim$0.8\arcsec), and thus it is not a spectrum of only the
contaminating diffuse component.  In any case, since both the target
spectrum and the diffuse component spectrum contain the pure SN and
pure diffuse light spectra, the subtraction should yield a pure SN
spectrum.

\begin{figure*}
\begin{center}
        \begin{minipage}[]{0.7\textwidth}
                \epsscale{1.0}
                \plotone{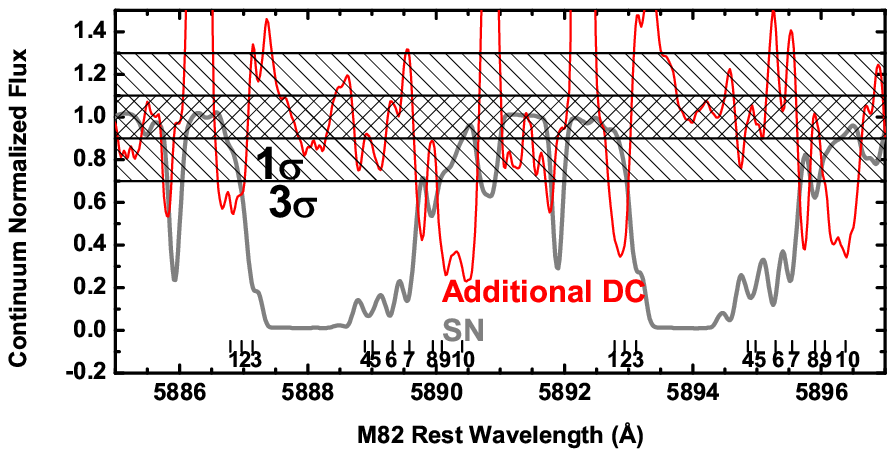}
        \end{minipage}
        \begin{minipage}[]{0.7\textwidth}
                \epsscale{1.0}
                \plotone{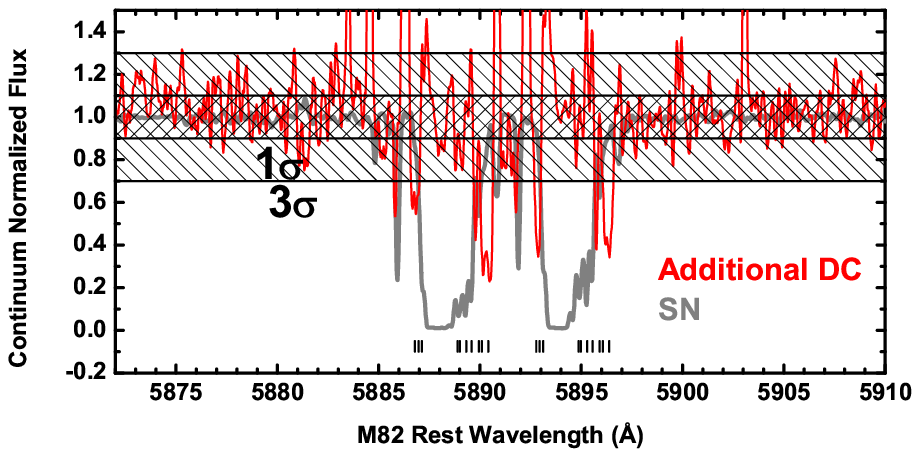}
        \end{minipage}
\end{center}
\caption{Difference between the $+255.1$~day total (diffuse-light contaminating) spectrum and the $+13.1$~day SN~2014J spectrum (red curve). The original
  spectrum is deconvolved into the SN component and this additional
  DC, and the resulting ratios (in the continuum flux) are $\sim$75\%
  and $\sim$25\%, respectively. This residual spectrum has been
  smoothed with a 7 pixel boxcar filter. Shown in gray is the
  \ion{Na}{1}~D profile for reference.  The gray hashed regions
  represent 1 and 3-$\sigma$ deviations between the SN and diffuse
  component. The lower panel shows an extended view in the dispersion
  direction to show the noise level of the resulting spectrum.
  \label{fig10}}
\end{figure*}

Figure~8 shows the resulting `pure' SN spectrum, normalized by the
continuum. While the S/N per pixel is reduced to
$\sim$15 -- 18 in the continuum around the \ion{Na}{1}~D (after the
binning of 7 pixels), we find that the resulting
contamination-subtracted line profiles are consistent with those in
the spectra from earlier phases. No significant variation between the
early- and late-time spectra is found any longer beyond $3\sigma$. This is
clearly seen in the blue and red wings of the saturated components
where the unsaturated components are present (i.e., components 1--3
in the blue and components 8--10 in the red).  In particular, the
spectral slopes in the blue and red parts of the saturated components
(blueward of component 1 and for the wavelength region around
component 10), originally flatter (deeper) than those in the earlier
phases, are now steeper (shallower) than the original (unsubtracted)
spectrum, and consistent with those at early phases. The facts that
this is confirmed irrespective of the model for the contaminating
diffuse light spectrum and that it is seen in both of the D1 and D2
lines strongly support this conclusion.  Therefore, we conclude that
the \ion{Na}{1}~D absorbing systems in the `pure' SN spectrum, i.e.,
those exactly along the line-of-sight to SN~2014J (here within the
projected radius of the size of the SN at this phase, i.e.,
$\sim$0.02~pc), did not change significantly even $+255.1$~days after
$B$-band maximum brightness.

This also supports the idea that `all' absorbing systems are dominated
by giant absorbing clouds in front of the SN site. In \S 3.2, we
mentioned that, strictly speaking, components 1 and 10 could
potentially come from the region localized to the SN position (i.e.,
`CSM'). If these components, originally weak in the early phases,
became saturated on +255.1 days, then it is possible that these
components are truly local to the SN without any contribution from
foreground systems. However, if this is correct, we should see these
components nearly saturated in our `contamination-subtracted'
spectrum, but this is indeed the opposite of the observations.
Therefore, this possibility is rejected, and we conclude that all
absorption systems are dominated by those in the foreground.

\begin{figure*}
\begin{center}
        \begin{minipage}[]{0.7\textwidth}
                \epsscale{1.0}
                \plotone{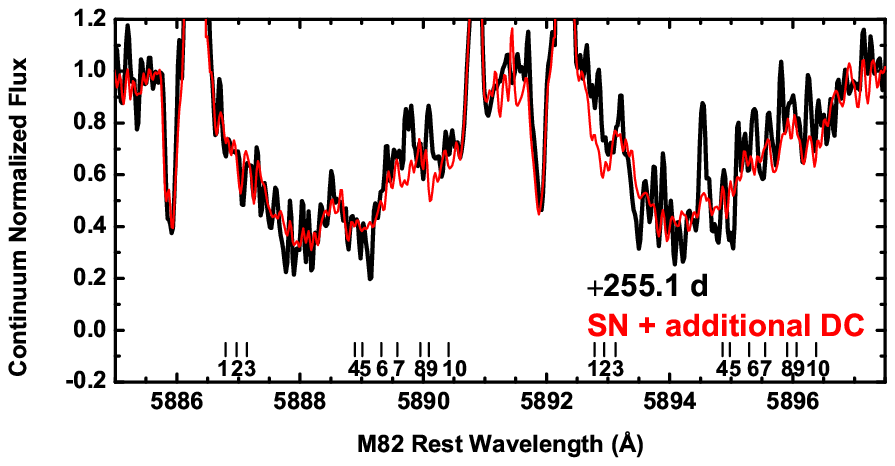}
        \end{minipage}
        \begin{minipage}[]{0.7\textwidth}
                \epsscale{1.0}
                \plotone{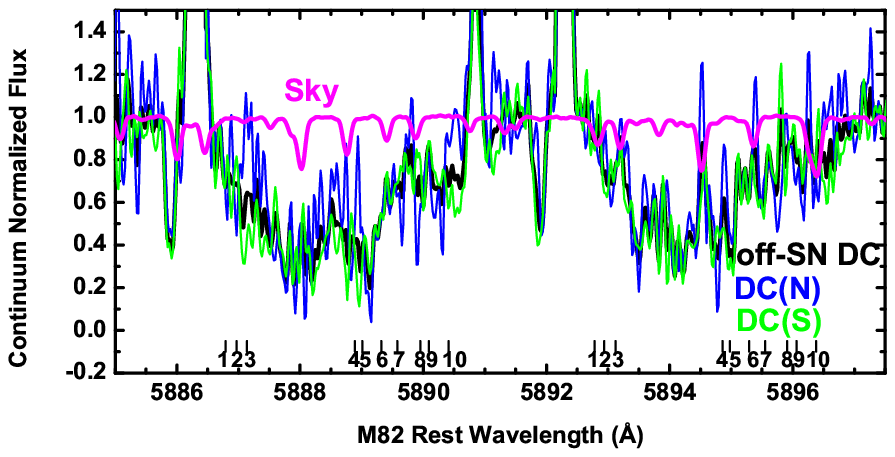}
        \end{minipage}
\end{center}
\caption{(Upper panel) Spectrum of the off-line-of-sight region,
  averaging the north and south regions (black: see Figure~1).  The
  red curve represents a reconstruction of this spectrum by combining
  the SN-only and DC components considered in Figure~10. The relative
  contributions to the continuum flux are 60\% from the SN component
  and 40\% from the pure (additional) diffuse component. Note that the
  relative contributions of these pure SN and DC components in this
  off-line-of-sight spectrum are different from those in the line-of-sight
  spectrum (see the Figure 10 caption). (Lower panel) North (blue) and
  South (green) background spectra as well as the averaged spectrum
  (black). The sky spectrum is also shown (magenta). \label{fig11}}
\end{figure*}

Another check on our conclusion can be provided by performing the flux
calibration to the contamination-subtracted pure SN spectrum. This is
shown in Figure~9 for the DC model 1. After the correction, the SN
shows no additional flux, and this pure SN spectrum provides a good
match to a spectrum of SN~2003du at a similar epoch. Also, the depths
of the absorption are now consistent with that seen in the early
phases, reaching to the zero-flux level in the saturated components.
All of these features support our interpretation.

\subsection{Spatial Variation in the Absorbing Systems}
The SN-site diffuse light globally shows a similarity to that directly
along the line-of-sight to SN~2014J, and therefore the bulk of the
absorbing systems are in the foreground region. However, we have not
yet quantitatively demonstrated that the absorbing systems for the
diffuse light and SN are exactly identical.  Given the lack of temporal
variation in the `pure' SN component, the apparent difference between
the early-time and late-time (diffuse light-contaminated) spectra
means that there must be a difference between the `pure' SN and the
diffuse components, otherwise the sum must be always identical.

We assume that the pure SN component in the late-time spectrum is
identical to that in the earlier phases, and subtract it from the
original, diffuse light-contaminated late-time spectrum (the `total'
spectrum in Figure~1). We note that this procedure also subtracts the
diffuse light that has an absorption-line spectrum identical to the
pure SN spectrum. Namely, if the SN and the diffuse light (within the
aperture of $\sim$40~pc in radius) show exactly the same absorption
systems, there should be no residuals. We note that this procedure is
identical to one examining the time variability of the absorption systems
toward an SN. However, given our conclusion that basically all of the
absorbing systems are located far from the SN in the foreground region
(having a physical scale of $\gsim$40~pc in radius), we regard any
residuals, if they exist, as originating in the `spatial' variation
(within $\sim$40~pc in the radius).

The result of this exercise is shown in Figure~10, which shows a spectrum
of `additional' absorbing components after the `pure' SN spectrum has
been subtracted from the total spectrum. Inversely, we find that the
relative contributions in the total light are $\sim$75\% (the pure SN
+ the identical diffuse component) and $\sim$25\% (the additional
diffuse component superposed on the continuum) in the continuum flux.
The relative contribution reflects the depth of the `saturated'
components in the total light, which is $\sim$25\%. Using this
information, we can infer the origin of the contaminating light either
in the region in front of the absorbing system (i.e., foreground) or
behind it (i.e., background): here, the contribution to the total
light within the whole aperture from the foreground light must be
$\sim 25$\%, and the remaining $\sim 75$\% being the sum of the SN light
and the diffuse light, must come from the region behind the absorbing
cloud(s).

Features in the derived residual spectrum can be interpreted as
follows. First, in most of the wavelength range across the
\ion{Na}{1}~D absorption, the difference is within $3\sigma$.
Therefore, the majority of the absorbing systems are identical
between the line-of-sight to the SN and the off-line-of-sight (beyond
$\sim$0.02~pc from the SN center in projected radius). Of course, we
cannot quantify the difference in terms of the column density for the
saturated components, but there is no hint that the column density in
the off-line-of-sight is significantly smaller than that along the
line-of-sight. This strengthens our conclusion that the absorbing
systems are produced by foreground materials located between the SN
and us, extending to the scale of $\sim$40~pc in the projected radius.
As for the unsaturated components, numbers 3--8 show residuals
within $3\sigma$, showing no significant difference between the SN and
the diffuse component. Therefore, these regions must not have
significant variation across the projected radius of $\sim$40~pc,
suggesting these components, together with the saturated components,
are attributed to a giant cloud of absorbing matter extending more
than 40~pc in the physical scale. Note that these are the systems
showing the largest absorption (therefore a low flux level), and thus
the S/N ratio is quite low. Indeed, there is a hint that components 3
(and blueward) and 7 show `negative' additional absorption, but given
the relatively low S/N ratio in this spectrum of the additional diffuse
component (Figure~10), this is just indicative.

There are some absorption systems that vary beyond $3\sigma$: the
unsaturated components 1 ($-160$ km/s relative to the M82 rest frame),
2 ($-152$ km/s), 9 ($+7$ km/s), and 10 ($+3$ km/s). There is also a
significant absorbing system at the wavelength between the components 7
($-19$ km/s) and 8 ($0$ km/s) where the SN line-of-sight does not show
significant absorption. In addition to the statistical significance,
these are all seen in both \ion{Na}{1}~D1 and D2, and therefore these
features must be real. These should be components whose column
densities vary depending on the line-of-sight, within a scale of
$\sim$40~pc in projected radius. It is interesting to see that the
components showing the spatial variation are the systems having weaker
absorption (thus smaller column densities) than those showing no
variation. This may suggest that the spatially smaller systems tend to
have lower column density, which might support our interpretation
since the larger systems are expected to have a larger extent along
the line-of-sight resulting in a larger column density.

\subsection{Further Support for the Interpretation}
If our interpretation is correct, it is expected that the spectrum of
the diffuse light originates from the region off the line-of-sight to
the SN (i.e., that extracted from the region at $\gsim$1.4\arcsec) can
be reproduced by a sum of the `pure-SN' component and the
`additional-DC' component, but with relative contributions different
from those applied to the total integrated spectrum including the SN
line-of-sight. For the total spectrum, the relative contributions are
75\% (the pure SN + the identical DC) and 25\% (the additional DC
superposed on the continuum) in the continuum flux.

Figure~11 shows that the off-line-of-sight spectrum is well reproduced
by a combination of the pure SN (including the diffuse component
identical to the SN absorbing systems) and the additional DC
components (including the continuum flux), but with different
contributions than those applied to the total integrated spectrum. The
diffuse spectrum in the surrounding region, as obtained by the DC
model 2, is representative of the spectra in the annuli surrounding
the SN at $\sim$1.4 -- 2\arcsec\ (i.e., $\sim$30 -- 40~pc in projected
radius). In this figure, the model spectrum is created by adding the
pure SN and the additional DC spectra with relative contributions to
the continuum flux of 60\% and 40\%, respectively. The contribution of
the additional diffuse component is increased in the surrounding
region.  This strengthens our conclusion that the absorbing systems
are all from the foreground big system (extending in the spatial scale
of $\gsim$80~pc in diameter). We emphasize that the additional DC was
created from the whole region extending to $\sim$2\arcsec\ of the
aperture where a large fraction of light is within $\lsim$1.4\arcsec.
Therefore there is no reason to expect that the off-line-of-sight
region ($\sim$1.4\arcsec\ -- 2\arcsec) spectrum is explained by the
combination of the pure SN and the additional DC at the SN position,
{\em if} this additional component is attributed to anything else
(e.g., time-varying absorption along the SN line-of-sight).

\begin{figure*}
\begin{center}
        \begin{minipage}[]{0.9\textwidth}
                \epsscale{1.0}
                \plotone{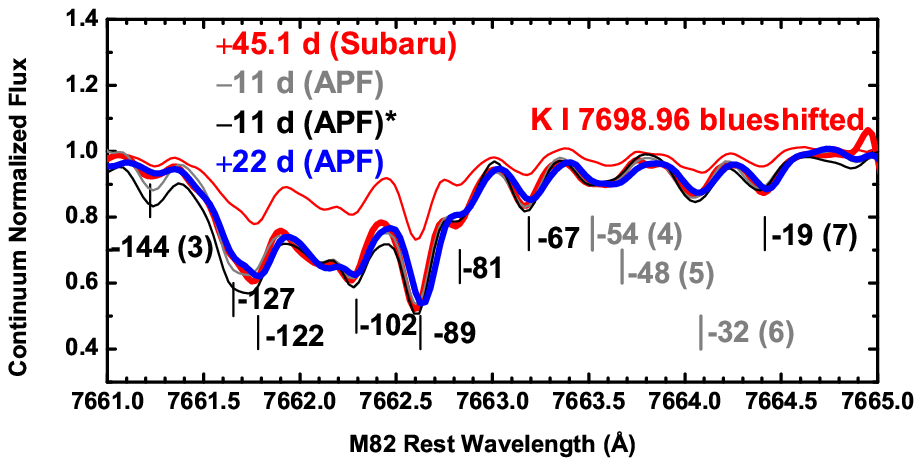}
        \end{minipage}
\end{center}
\caption{Continuum-normalized spectrum of SN~2014J at $+45.1$~days
  relative to $B$-band maximum brightness, shown for the \ion{K}{1}
  doublet ($7664.90$\AA\ in thick-red and $7698.96$\AA\ in thin-red).
  The red ($7698.96$\AA) component is artificially blueshifted in
  wavelength, so that it is shown in the wavelength scale of the blue
  ($7664.90$\AA) component. The \ion{K}{1} components identified by
  \citet{graham2014} are marked with black vertical lines, while
  others corresponding to the \ion{Na}{1}~D components are marked with
  gray vertical lines. The numbers denote the blueshift in
  km~s$^{-1}$, while those in parenthesis indicate the corresponding
  \ion{Na}{1}~D component numbers. The continuum-normalized spectra of
  the first and last spectra obtained by the Automated Planet Finder
  (APF) at Lick Observatory \citep{graham2014} are shown (one at
  $-11$~days in thin-gray and at $+22$~days in thick-blue), where the
  fringing pattern (see the main text) was reconstructed using the
  same spectrum. Also shown is the result adopting a second fringe
  correction model for the $-11$~day spectrum (thin-black). \label{fig12}}
\end{figure*}

The same argument can be rephrased in a slightly different way. The
contribution from the background light (i.e., that from the region
behind the absorbing systems) reflects the depth of the `saturated
component', which is $\sim$40\%. This means that in the off-SN region,
the contribution of the light from the foreground (in front of the
absorbing cloud) is $\sim$40\% with the rest coming from the region
behind the absorbing cloud. As the SN light contamination must be
negligible along the line-of-sight for this off-SN region to a first
approximation, this means that generally near the site of SN~2014J,
$\sim$60\% of the light is a true background (behind the absorbing
systems) and $\sim$40\% is a foreground (in front of the absorbing
systems).

This is consistent with the contribution of the foreground light
reaching $\sim$30\% in the total light within the aperture (i.e., the
SN and DC), as inferred from the flux of the `saturated' components of
\ion{Na}{1}~D. The SN must be fully behind the absorbing systems, and
the SN contribution within the aperture is $\sim$25\% in the continuum
around \ion{Na}{1}~D (see Figure~4). Then the contribution from the
foreground to the total light should be $\sim$(1 -- 0.25) $\times$
$0.4 \approx 0.3$, i.e., $\sim$30\%.

Because of the image rotation during the exposure, we lose information
on the angular distribution of the incoming light around the center of
the pointing to some extent (Figure~1). Given the relatively low flux
in the off-SN diffuse light (i.e., DC model 2), we have created the
off-SN diffuse light spectrum by averaging those on both side of
the SN position. However, it is still possible to use some spatial
information. Figure~11 shows the spectra of the two spatially distinct
off-SN diffuse regions (N, north, and S, south) separately. The
spectra of these two regions generally match one another quite well,
supporting the idea that the absorbing systems are from the foreground
with a physical scale of $\gsim$40~pc in radius. However, there are
marginal differences between the two, especially near the unsaturated
component 3 (slightly blueward) and the red part of component 10.  The
former is not seen in \ion{Na}{1}~D1, but that wavelength overlaps
with a relatively strong telluric absorption. The former corresponds
to the `additional DC' where the `negative' absorption is marginally
seen, and the latter corresponds to the wavelength where the
additional DC absorption is detected (Figure~10, upper panel).  These
indicate that these components show relatively small-scale
fluctuations within $\sim$30~pc.

\subsection{\ion{K}{1} and DIBs}

Here we examine the claim by \citet{graham2014} of variability in the
\ion{K}{1} profile corresponding to the unsaturated \ion{Na}{1}~D
component 3 ($-144$~km~s$^{-1}$) and part of the saturated component
($-127$~km~s$^{-1}$). We have only one spectrum covering the
wavelength of \ion{K}{1}, but this spectrum is of very high quality
and was taken on $+45.1$~days, about 23~days later than the last
spectrum presented by \citet{graham2014}. In addition, we note that
the spectra taken by the Automated Planet Finder (APF) at Lick
Observatory \citep{graham2014} to some extent suffer from a fringing
pattern in defining the continuum, and thus we want to provide a check
with data taken by another instrument. In doing this, we tried to
correct for the fringing pattern in the spectra of \citet{graham2014}
using our Subaru/HDS spectrum on $+45.1$~days, which is free from such
a fringing pattern. Details are described in Appendix~B.

Figure~12 shows the comparison, where our spectrum on $+45.1$~days is
compared with the APF spectra at $-11$~days and $+22$~days after
correcting for the fringing pattern. First, even without the
correction for the fringing pattern, we find that the features in
\ion{K}{1} on $+45.1$~days are generally in good agreement with those
at $+22$~days presented by \citet{graham2014}. The absorption depth of
the component at $-144$~km~s$^{-1}$ is no deeper than $0.9$, and the
depth of the component at $-127$~km~s$^{-1}$ is significantly
shallower than that at $-122$~km~s$^{-1}$. In addition, the same
features are confirmed in the red component of the \ion{K}{1} doublet
in our spectrum. This supports the idea that the features of \ion{K}{1} in the
APF spectrum at $+22$~days as reported by \citet{graham2014} are real,
and indicates that the features did not show significant variability
between $+22$ and $+45.1$~days.

After removing the fringing pattern in the APF spectra, we find an
excellent match between the spectra at $+22$~days and $+45.1$~days. No
significant evolution is found between the two epochs. The spectra at
$-11$ and $+22$~days, with our best effort to correct for the fringe
pattern using our Subaru spectrum, reproduce the time evolution found
by \citet{graham2014}: at the later epoch, the component at
$-144$~km~s$^{-1}$ became shallower, and the component at
$-127$~km~s$^{-1}$ became shallower as compared to the one at
$-122$~km~s$^{-1}$. Thus, we have confirmed the main findings of
\citet{graham2014} on the variability in \ion{K}{1}. In addition, our
new spectrum shows that no evolution is seen from $+22$~days to
$+45.1$~days.

A question then is how the \ion{K}{1} variability is consistent with
our finding that all the absorbing systems seen in \ion{Na}{1}~D are
caused by foreground absorbers extended at a scale of $\gsim$80~pc,
with a significant fluctuation in the column density for some
relatively weak unsaturated components at a scale of $\sim$20~pc.
\citet{graham2014} estimated the distance of the foreground absorbing
material, showing the \ion{K}{1} variability as being at 3--9~pc from
the SN site. Their finding is consistent with our conclusion if the
gas is relatively close to the SN but spatially extended in the plane of the sky.
It is interesting that we do see a hint of spatial variation in the
\ion{Na}{1}~D absorbing systems within $\sim$20 -- 30~pc corresponding
to those systems showing the temporal variability in \ion{K}{1} (Figures 10
and 11). This supports the origin of this component at a scale of
$\sim$10 -- 20~pc from the SN in the foreground region.  It is notable
that the closest material to the SN also has the highest blueshift.

\begin{figure*}
\begin{center}
        \begin{minipage}[]{0.3\textwidth}
                \epsscale{1.2}
                \plotone{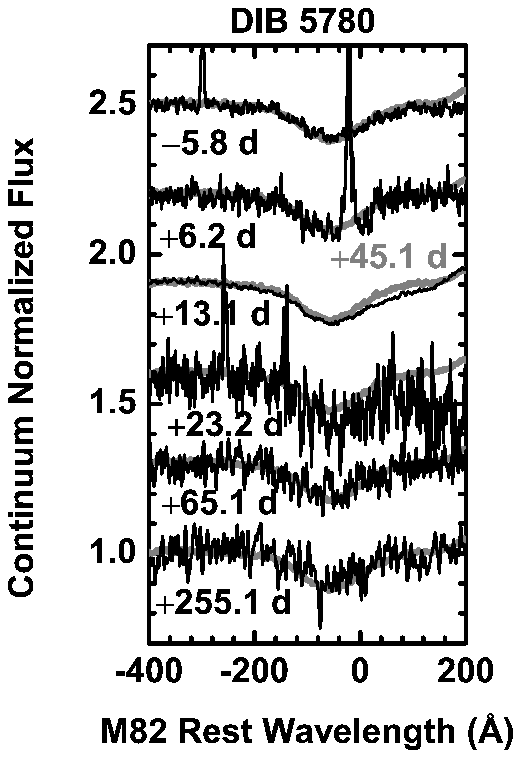}
        \end{minipage}
        \begin{minipage}[]{0.3\textwidth}
                \epsscale{1.2}
                \plotone{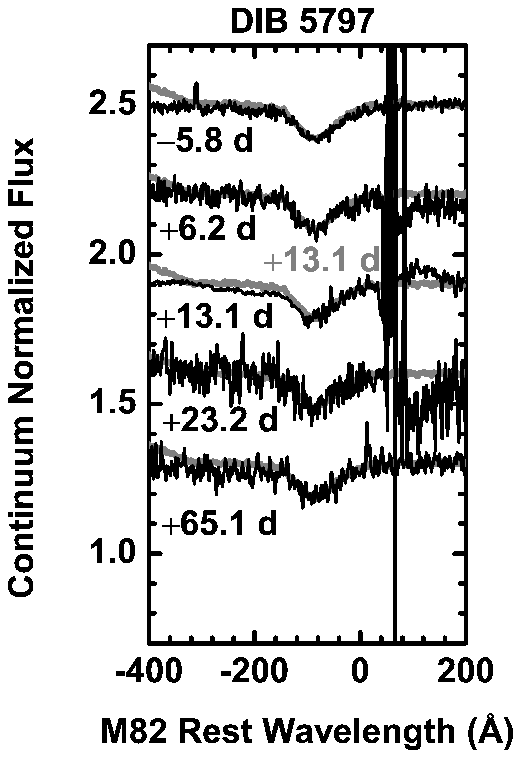}
        \end{minipage}
        \begin{minipage}[]{0.3\textwidth}
                \epsscale{1.2}
                \plotone{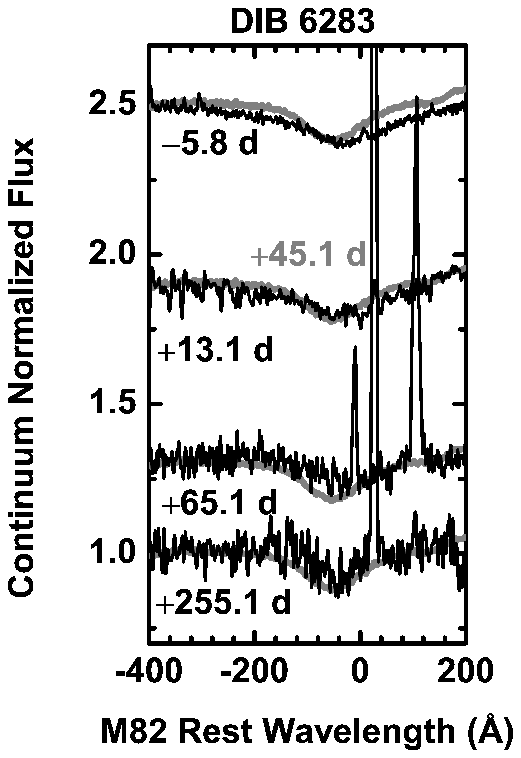}
        \end{minipage}
\end{center}
\caption{Spectra focusing on DIB features. The 5797~\AA\ DIB in the
  $+255.1$~day spectrum falls in a CCD gap. The 6823~\AA\ DIB feature
  is too noisy in the $+6.2$ and $23.2$~day spectra and is not
  displayed here. \label{fig13}}
\end{figure*}

No significant variation of DIBs has been reported between $-10$ and
$+45$~days \citep[for DIBs 5780 and 5797: ][]{foley2014} and between
$-10$ and $+22$~days \citep[DIBs 5780, 5797, 6196, 6283, 6613:
][]{graham2014}. Figure~13 shows the evolution of DIBs in our spectra,
up to $+255.1$~days for DIBs 5780 and 6283, and up to $+65.1$~days for
DIB 5797. While the analysis of DIBs is limited by the S/N ratio and
the accuracy of the continuum subtraction, no significant variation is
seen for these features up to $+255.1$~days. This is consistent with
the idea that the DIBs mainly originate in the ISM \citep{vanloon2009,vanloon2013}, and thus the
existence and strengths of DIBs can be used to trace the ISM
components and extinction toward SNe \citep{phillips2013}.
Interestingly, this result provides a good contrast to variation in
some DIBs reported for broad-lined SN~Ic~2012ap associated with the
death of a massive star \citep{milisavljevic2014}. While the
equivalent widths of DIBs are comparable for SNe~2014J and 2012ap, no
variation is seen for SN~2014J even at $+255.1$~days (note that this
study presents the longest span so far for any SNe for which such an
evolution in DIBs has been tested). It highlights the difference
between SNe~Ia and core-collapse SNe, and may shed light on the still
enigmatic origin(s) of DIBs.

\section{Conclusions and Discussion}

We have discovered apparent `variability' of some 
the \ion{Na}{1}~D absorption systems toward SN~2014J. We however conclude
that this variability does not reflect intrinsic time variability of
the absorbing system along the line-of-sight toward the SN, but
rather reflects the contamination of an unassociated diffuse light
whose contribution was negligible when the SN was bright. Indeed, we
note that the similar behavior might have been already seen previously
in the spectrum of SN~2007le at $+90$~days \citep{simon2009}; which
might have been caused by bad seeing in the observation leading to a
contamination of an off-SN site diffuse light to the true SN spectrum
in that case. The diffuse light contamination, usually an obstacle in
isolating the properties of the line-of-sight information toward an
SN, however turns out to be very powerful in identifying the origin of
the strong and complicated absorbing systems seen in SN~2014J.
Specifically, this study presents the first case where all absorbing
systems toward an individual SN are unambiguously deconvolved into
ISM and CSM components (resulting in no CSM components for SN~2014J)
-- Extraction of the CSM components has been successful either using
time variability for individual objects \citep[e.g.,][]{patat2007,
  simon2009} or using a large number of SNe for a statistical sample
\citep[e.g.,][]{sternberg2011}. However, both routes ineitably lack the
identification of the origin(s) of the non-variable components for
individual SNe.

\begin{figure*}
\begin{center}
        \begin{minipage}[]{0.9\textwidth}
                \epsscale{1.0}
                \plotone{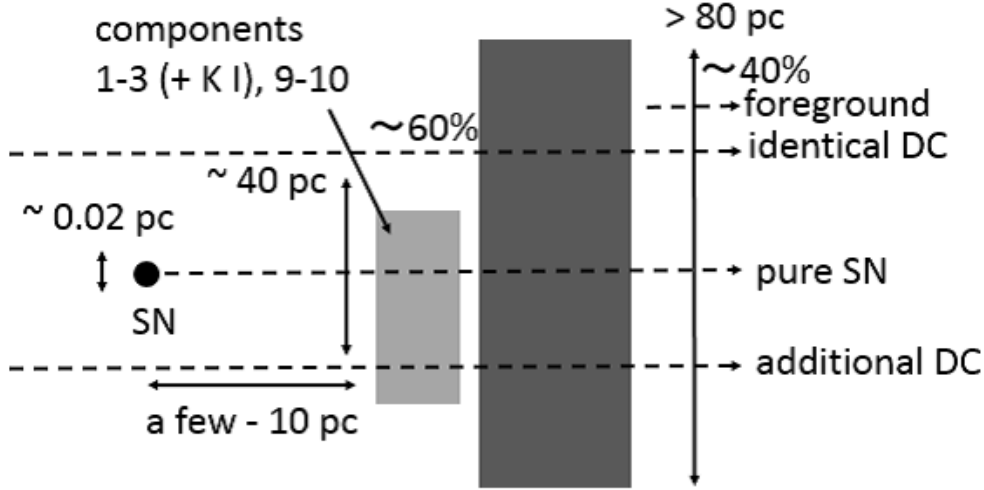}
        \end{minipage}
\end{center}
\caption{Sketch of the \ion{Na}{1}~D absorbing systems toward
  SN~2014J. \label{fig14}}
\end{figure*}

The inferred structure of the absorbing systems is schematically shown
in Figure~14. All absorbing systems seen toward SN~2014J should be
located in a foreground region far from the SN site since they show
the depths of the absorption clearly exceeding the value expected 
for a system localized along the SN line-of-sight (\S 3.2 and Figure~7). 
That is, the absorbing spectrum of the contaminating light is generally
identical to that intrinsically toward the SN (i.e., within
$\sim$0.02~pc of a pencil beam toward the SN). The physical scale of
the absorbing system must be (generally) greater than $\sim$80~pc because
the spectrum does show generally identical absorbing systems at this
scale. The `saturated' components show a non-zero flux at a level of
$\sim$30\% relative to the continuum, and it is likely that this is
caused by a contamination from the foreground region with respect to
the location of the bulk absorbing materials. The intrinsic (without
SN) contribution of the foreground light reaches $\sim$40\% near the
SN site.

Identifying the diffuse light contamination, we have managed to
subtract this from the total light and have extracted a `pure'
absorbing spectrum toward the SN. This pure line-of-sight spectrum
does not show significant change in the strength of the absorbing
systems as compared to the spectra at earlier phases. Namely, the
intrinsic line-of-sight absorbing systems within a narrow
line-of-sight toward the SN did not change even at $+255.1$~days.
This is in accordance with our conclusion that all of the visible
absorbing systems are located far from the SN site and not at close to
the SN (i.e., not CSM).

At the same time, we see small, but significant, differences in the
absorbing systems in the late-time spectrum (SN plus the diffuse
light) and in the earlier phases. The difference is seen in some
unsaturated components. Since there is no evidence for temporal
variability along the SN line-of-sight, this variation must come from
the `spatial' variation at a scale of $\lsim$40~pc in projected
radius. This is further supported by the fact that this difference
strengthens at larger radii.

Another appealing possibility, not mentioned in \S 3, is that this
apparent variation in the spatial dimension might reflect a light
echo, where the SN light is scattered off dust sheets to us.  A
projected light echo position can show a superluminal motion and thus
can extend to the projected distance of $\gsim$20~pc. If this is true
it would indicate an angle-dependent inhomogeneity in the
absorption systems, but we regard this possibility as highly unlikely.
\citet{crotts2014} reported that diffuse components appeared at about
$+213$~days in the {\it HST} image, associating them with the SN light
echo.  The components are however much fainter than the SN itself, and
thus this would not explain the apparent difference in the
absorption-line spectra in the earlier epochs and that at
$+255.1$~days. Our flux-calibrated spectrum at $+255.1$~days shows a
normal SN~Ia nebular spectrum, showing no clear signature of a light
echo. We note, however, that the continuum of the SN-site diffuse
light is relatively blue, and it may indicate that there could
possibly be some contamination from an SN light echo there. While the
contamination would not be at a level to change our conclusions,
investigating the possible unresolved light echo component itself is
an interesting topic and continual follow-up observations of SN~2014J
are highly encouraged.

We confirmed the finding of \citet{graham2014} for the variability of
\ion{K}{1}, but also extended the analysis to a later epoch. The
nature of variable \ion{K}{1} components, at $-144$~km~s$^{-1}$ and
$-127$~km~s$^{-1}$ (with respect to the M82 rest frame), in the
$+45.1$~day spectrum is consistent with that at $+23$~days, indicating
that those components have stabilized after $+23$~days.  Given the
conclusion by \citet{graham2014} that the system is likely located at
a distance of 3--9~pc from the SN, this does not conflict with our
findings regarding the \ion{Na}{1}~D absorption systems -- they are
both in the foreground region. DIBs did not show significant evolution
from $-5.8$~days to $+255.1$~days, further supporting the location of
the absorbing system being far from the SN site.

The derived configuration is consistent with the {\it HST}
high-resolution pre-SN image (see Fig. 1). SN~2014J is in the region
where a dark lane is seen, and the region typically shows a
fluctuation in the surface brightness at a scale of $\sim$20~pc.

While we do not constrain the line-of-sight distance between the SN
site and the absorbing materials \citep[e.g., suggested to be 3--9~pc
for the highest-velocity clouds as determined from the \ion{K}{1}
variability:][]{graham2014}, we do constrain the projected size of the
systems as being at least $\sim$40~pc. Therefore, if the absorbing
material is related to pre-SN activity, then it had to affect a large
distance ($\sim$40~pc) rather than a few pc. This implies that it is
unlikely that the absorbing material is directly related to
progenitor activity.

To quantify this point, we divide the question into two sub-parts.
The first question is whether there is any possibility that the absorbing
materials at this distance can be explained by material ejected from
the SN progenitor system -- the answer is no. If the SN progenitor had
affected a region of this size ($\sim$40~pc in radius), assuming
spherical symmetry, the swept-up materials from the ISM would have to be as
large as $\sim$6000~$M_{\odot} (n_{\rm ISM}/ 1 \ {\rm cm}^{-3})$,
greatly exceeding `CSM' material ejected from any proposed systems for
SNe~Ia. The next question is whether there is a chance that this region was
affected by the progenitor system activity -- the answer is that it is possible
but not likely, as highlighted below when considering cases expected
from the SD scenario.

If this material has been accelerated by a wind from a red-giant
companion star with $10^{-6} M_{\odot}$~yr$^{-1}$ and a wind velocity
of $\sim$30~km~s$^{-1}$ (i.e., an extreme case of the symbiotic path
in the SD scenario), namely at a power of about $3 \times
10^{32}$~erg~s$^{-1}$, it would take about $2 \times 10^{7}$ years for
the swept-up materials to reach $\sim$40~pc. In this case the expected
expansion velocity should have decreased to a few km~s$^{-1}$, even if
radiation energy loss from the expanding shell is ignored. This
suggests that the ISM expansion caused by the hypothesized companion
giant wind would not reach this large spatial extent. In a symbiotic
case within the SD scenario, the ISM-shell expansion might be further
energized by recurrent nova activities. If the system emitted a
kinetic energy of $\sim 10^{43}$~erg in each eruption with a
recurrence time of $\sim$10~years, then the averaged kinetic input to
the surrounding ISM would be about $3 \times 10^{34}$~erg~s$^{-1}$.
In this case, the swept-up ISM would take about $5 \times 10^{6}$~yrs
to reach $\sim$40~pc. The resulting expansion velocity would be
$\sim$10~km~s$^{-1}$. This possibility is not rejected, but then the
bulk kinematics seen in the absorbing systems (reaching to about
$-150$~km~s$^{-1}$ for the component showing the variability in
\ion{K}{1}) must be attributed to the ISM kinematics unrelated to the
shell expansion.  At this point, the line velocity cannot be explained
by a wind. The largest impact on the gas surrounding the SD progenitor
system would be expected by an optically thick accretion wind, which
would have a kinetic power four orders of magnitude larger than the
red-giant companion wind. In this case it would take about
$10^{6}$~yrs for the ISM shell to reach $\sim$40~pc, and the resulting
expansion velocity would be $\sim$50~km~s$^{-1}$. This is probably the
only scenario, within the hypothesis that the absorbing systems are
directly related to the progenitor, that could account for the
kinematics of the blueshifted absorbing systems. However, in this case,
we would expect to see a supersoft X-ray source in pre-explosion
images, while none was detected \citep{nielsen2014}.

To push the above arguments further, we point out that it is generally
difficult to explain the kinematics of the absorbing system toward
SN~2014J as merely coming from the SN progenitor system, irrespective of
any assumptions about the system. This requires an excessively large
kinetic power input from the progenitor system. It is naturally
expected that such an energy input may well be followed by
electromagnetic radiation at some wavelength, while there were no
detections of such a source in pre-explosion images \citep{goobar2014,
  nielsen2014, kelly2014}. Therefore, the absorbing systems towards
SN~2014J should have no direct relation to the SN progenitor system
activity.

The non-detection of radio and X-ray signals from the putative SN
ejecta-CSM interaction places a strong constraint on the amount of the
CSM around the SN within $\sim$0.01~pc \citep{perez2014,
  margutti2014}. Properties of the polarization, especially the
non-detection of time variability, imply that the source of the
polarization is in the ISM \citep[although, see
\citealt{hoang2015}]{kawabata2014, patat2014}. No NIR/IR echo has been
identified for SN~2014J, which places a strong upper limit on the
amount of CSM \citep{johansson2014} \citep[see also][for a light echo
model and constraints for SNe~Ia in general]{maeda2014}.  Finally, the
present study shows that all observed narrow-line absorptions likely
originate from material in the ISM.  Strictly speaking, our analysis
does not reject the possible existence of CSM around SN~2014J, but it 
does reject any association of the observed (narrow) absorbing systems
with such CSM. There are indeed suggestions that the extinction curve
toward SN~2014J is better explained by a `multiple scattering model'
\citep{goobar2008, amanullah2014} than the ISM dust model, and that
the non-standard extinction curve toward SN~2014J could be explained
by a combination of ISM and CSM components \citep{foley2014}
\citep[but see][]{amanullah2015}. We have not yet arrived at a unified
explanation for all of these constraints, but our result adds to
another strong constraint on the existence of CSM.

To our knowledge, this is the first case where the observed
\ion{Na}{1}~D absorbing systems for an individual SN~Ia are
unambiguously associated with either the ISM or CSM, beyond arguments
using statistics and correlation between different observables
\citep[see, e.g.,][]{phillips2013}. It is interesting to note that
SN~2014J is the best example where the blueshifted (in the host-galaxy
rest frame) absorbing systems are clearly detected. Of course the
Doppler shift in the absorbing systems for a single object cannot be
(and has not been) used to associate the CSM materials with the observed
absorbing systems given the unknown bulk kinematics of the foreground
ISM \citep{sternberg2011}, but the fact that this most outstanding
case turns out to be unrelated to CSM may have important implications.
The kind of analysis in this paper provides a strong test for an
association of the blueshifted \ion{Na}{1}~D lines and the CSM for
individual SNe~Ia. To push this further, one may want to expand the
analysis to a sample of SNe~Ia where the existence of CSM components
in the observed absorbing systems is clearly limited for individual
SNe even without time variability. Our approach was possible for
SN~2014J because of its proximity (such cases are likely only 
once in a couple of decades), and
such an analysis would be impossible for other SNe~Ia even with
8m-class telescopes.  This, however, would be overcome in the future in 
an era of 30m-class telescopes, with which the horizon for a similar
analysis would expand to SNe~Ia at $\sim$15~Mpc or even further with
an adaptive optics system operating at optical wavelengths to
improve the S/N for extragalactic point sources. The targets can
be any nearby SNe within the limiting distance including those already
discovered, since the main part of the analysis requires only the
spectrum of the diffuse light around an SN, leaving a reasonable
number of targets for such a study. This would not only test the
progenitor systems of SNe~Ia, but it could be extremely powerful in 
determining the properties of the ISM, e.g., molecular content and origin(s) of
DIBs.

Indeed, a further follow-up at the site of SN~2014J in a
high-resolution mode will be valuable not only to confirm the present
results but also to obtain further insights into the origin of the
narrow-line absorbing systems. In the analysis presented here, the
off-SN diffuse region is still somewhat contaminated by the SN, which
complicates the interpretation of the pure diffuse components. Once
the SN has faded away, we should be able to obtain a pure diffuse
component spectrum. Furthermore, multiple-pointing observations can
give us the spatial distribution of different absorbing components
directly.

\acknowledgements

The authors thank the staff at the Subaru Telescope, Okayama
observatory, and Gunma Astronomical observatory for their excellent
support of the observations, especially for the flexible arrangement
and the support of the ToO observations. We are especially grateful to
H.\ Izumiura and E.\ Kambe for their help in obtaining the ToO
observations with the OAO 1.88-m telescope. The authors thank M.\
Graham for kindly providing the APF spectra of SN~2014J and for
stimulating discussion on the \ion{K}{1} absorbing systems. This paper
uses data based on observations made with the NASA/ESA Hubble Space
Telescope, obtained from the data archive at the Space Telescope
Science Institute. STScI is operated by the Association of
Universities for Research in Astronomy, Inc.\ under NASA contract
NAS~5-26555. The work by K.M.\ is partly supported by World Premier
International Research Center Initiative (WPI Initiative), MEXT,
Japan. The work has been supported by Japan Society for the Promotion
of Science (JSPS) KAKENHI Grant 26800100 (K.M.), 15H02075, 15H00788
(M.T.), 23224004, 26400222 (K.N.) and by JSPS Open Partnership 
Bilateral Joint Research Project
between Japan and Chile (K.M.).  R.J.F.\ gratefully acknowledges
support from NSF grant AST-1518052 and the Alfred P.\ Sloan
Foundation. K.N.\ is supported by WPI Initiative, MEXT, Japan. 

\appendix

\section{Analyses of Individual Exposures}

\begin{figure*}
\begin{center}
        \begin{minipage}[]{0.45\textwidth}
                \epsscale{0.9}
                \plotone{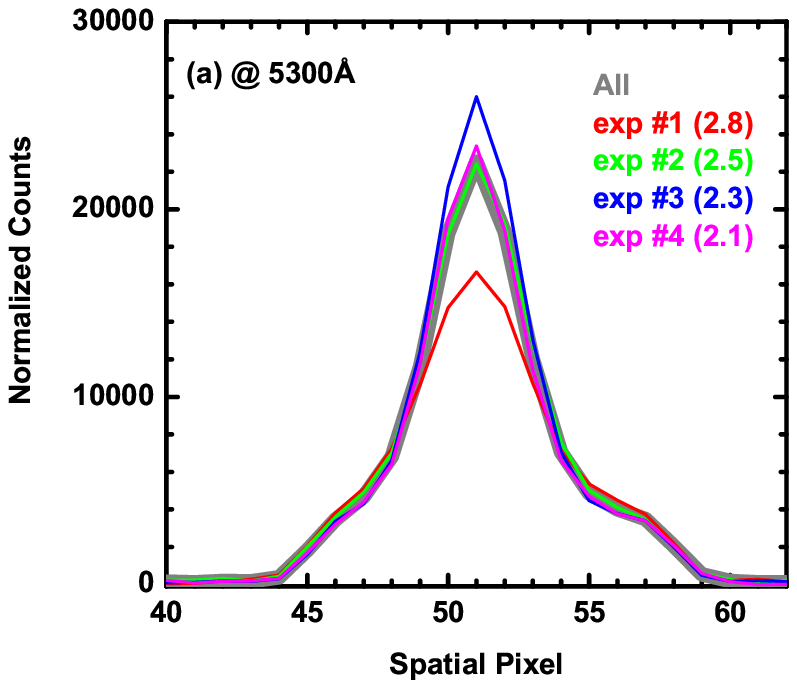}
        \end{minipage}
        \begin{minipage}[]{0.45\textwidth}
                \epsscale{0.9}
                \plotone{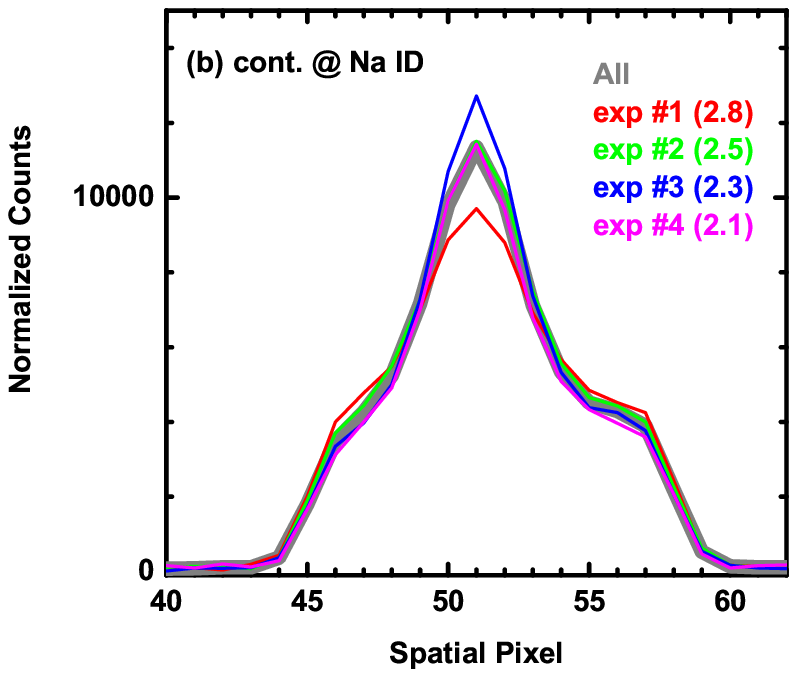}
        \end{minipage}
        \begin{minipage}[]{0.45\textwidth}
                \epsscale{0.9}
                \plotone{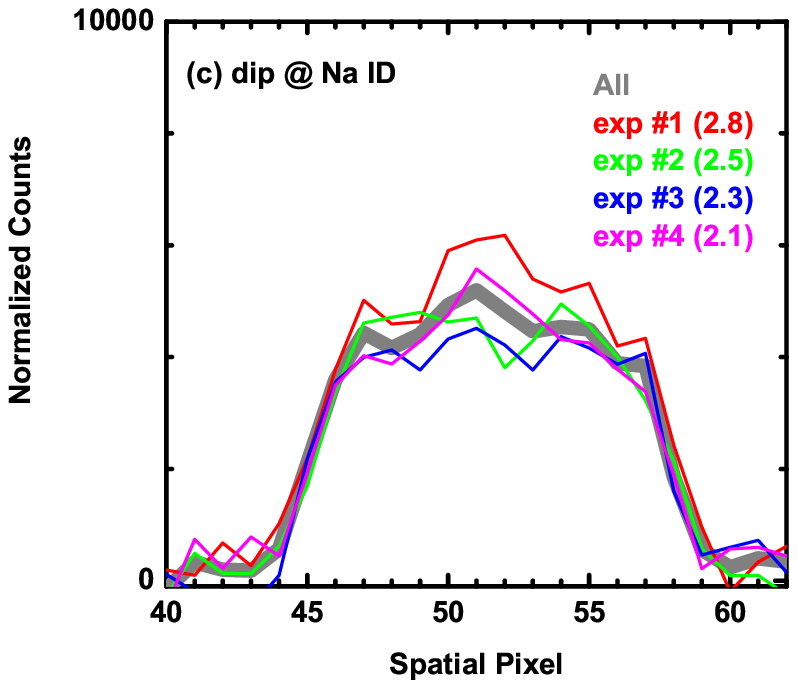}
        \end{minipage}
        \begin{minipage}[]{0.45\textwidth}
                \epsscale{0.9}
                \plotone{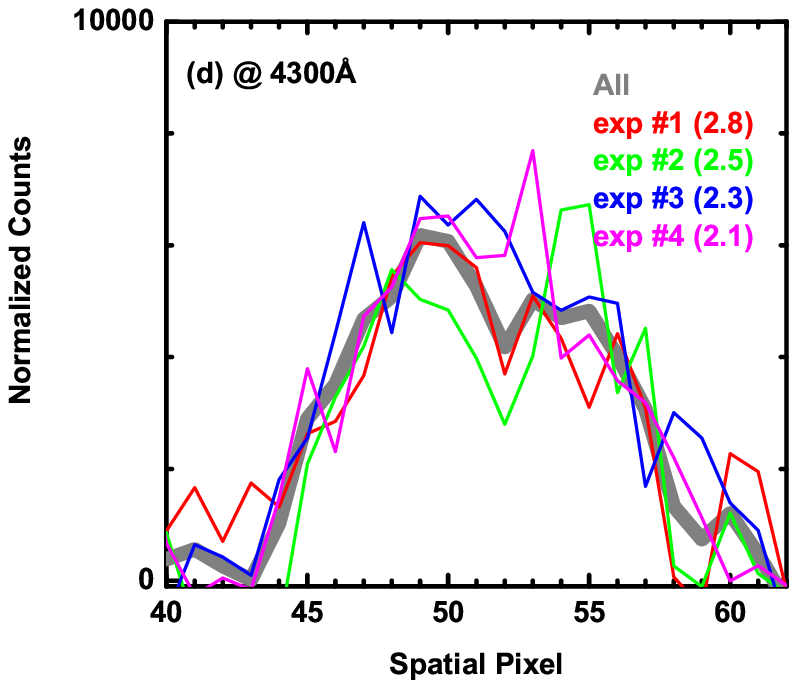}
        \end{minipage}
\end{center}
\caption{(Figure~A1): Spatial profiles of each individual exposure of
  the $+255.1$~day SN~2014J spectrum for different wavelength regions:
  (a) at $\sim$5300~\AA, (b) a continuum around \ion{Na}{1}~D, (c) a
  dip at \ion{Na}{1}~D, and (d) at $\sim$4300~\AA. In each panel, the
  spatial profile extracted from the combined 2D spectrum is shown in
  gray, while the profiles extracted from individual exposures are
  shown in other colors.  The airmass for each exposure is shown in
  parentheses in the labels. \label{figapp1}}
\end{figure*}

\begin{figure*}
\begin{center}
        \begin{minipage}[]{0.45\textwidth}
                \epsscale{1.0}
                \plotone{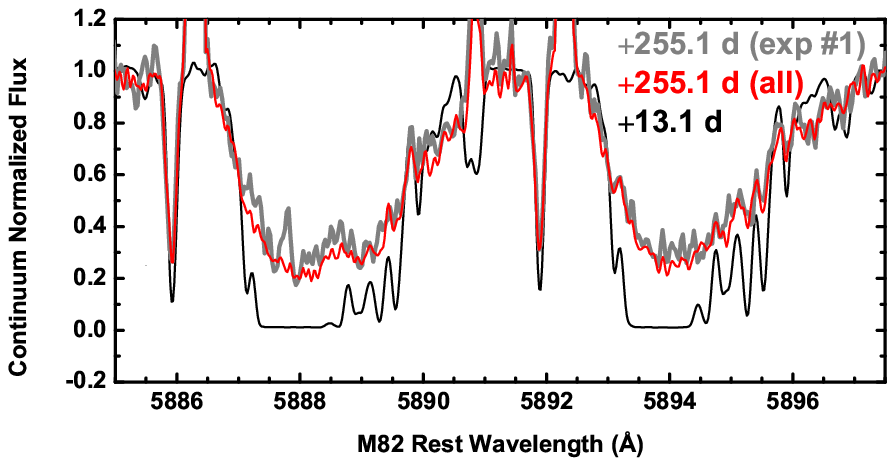}
        \end{minipage}
        \begin{minipage}[]{0.45\textwidth}
                \epsscale{1.0}
                \plotone{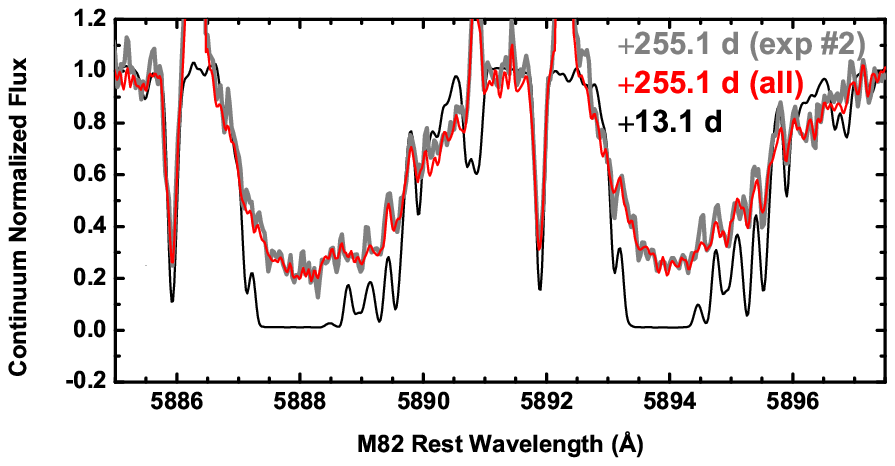}
        \end{minipage}
        \begin{minipage}[]{0.45\textwidth}
                \epsscale{1.0}
                \plotone{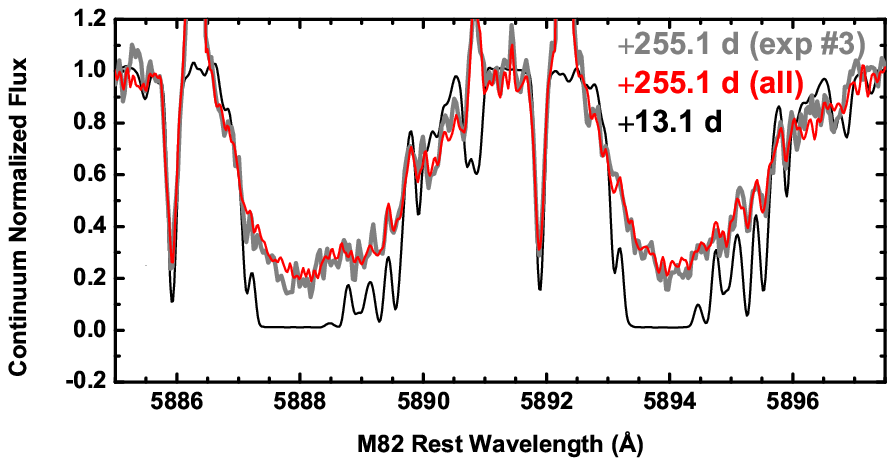}
        \end{minipage}
        \begin{minipage}[]{0.45\textwidth}
                \epsscale{1.0}
                \plotone{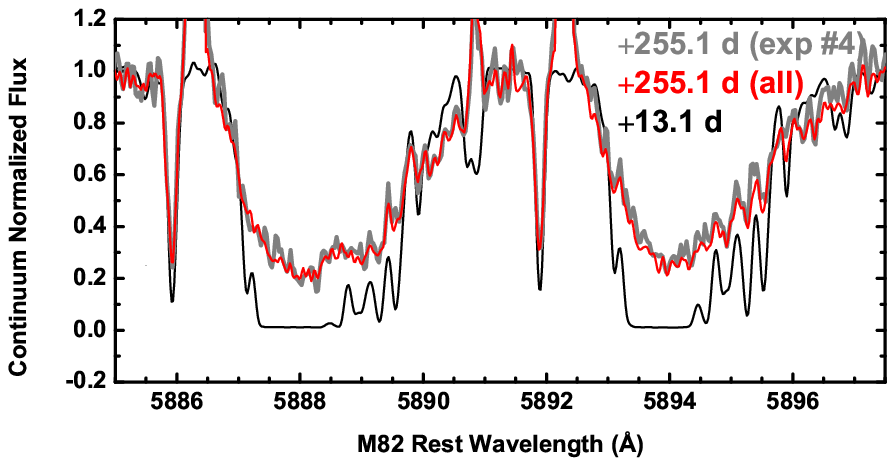}
        \end{minipage}
\end{center}
\caption{(Figure~A2): The $+255.1$~day spectrum of SN~2014J without
  subtracting the diffuse light, shown for each individual exposure
  (gray). For comparison, the spectrum extracted from the combined 2D
  spectrum is shown in red, as well as the $+13.1$~day spectrum in
  black. \label{figapp2}}
\end{figure*}

Our observation at $+255.1$~days was performed at high airmass,
ranging from $\sim$2.8 in the first exposure to $\sim$2.1 in the
fourth exposure. As such, we want to test whether this high airmass and its
evolution as a function of time could result in an unusually large
seeing and/or temporal variability of FWHM depending on the airmass to
a level that could affect our conclusions.

As shown in Figure~4, a point source (i.e., the SN) was clearly
detected with FWHM of $\sim$0.8\arcsec. Figure~A1 shows the spatial
profiles of the incoming light extracted from the individual 2D images
for each exposure. Indeed, there is a slight evolution in the FWHM of
the point source component -- it was $\sim$1\arcsec\ in the first
exposure but stayed at $\sim$0.8\arcsec\ in the remaining exposures.
Nevertheless, the line profiles at different wavelengths are consistent
between different exposures, and the change in the seeing is too small
to affect the observational features discussed in the main text.

Figures~A2 and A3 show the continuum-normalized total light (A2) and
DC-subtracted (A3) spectra of SN~2014J at $+255.1$~days. These are the
same as in Figures~5 and 8, respectively, but extracted separately for
each exposure. The individual exposures are consistent with each other
and with the spectrum from the combined 2D image. From these analyses,
we conclude that the high airmass did not affect any of our
conclusions.

\begin{figure*}
\begin{center}
        \begin{minipage}[]{0.45\textwidth}
                \epsscale{1.0}
                \plotone{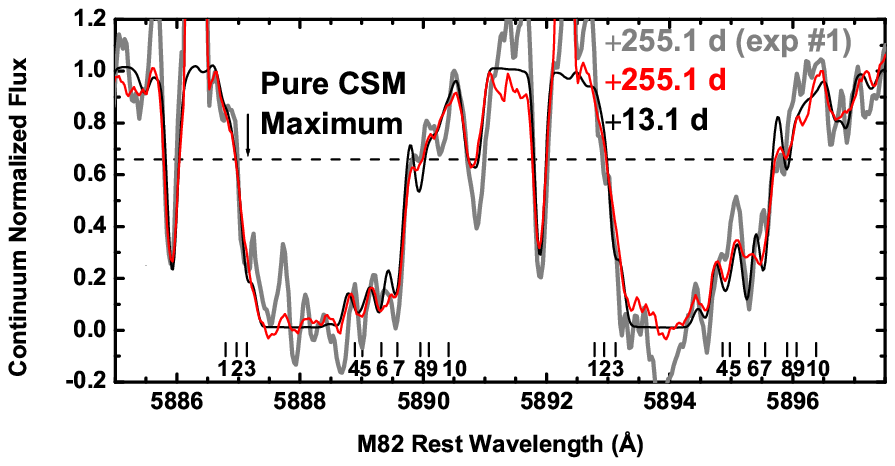}
        \end{minipage}
        \begin{minipage}[]{0.45\textwidth}
                \epsscale{1.0}
                \plotone{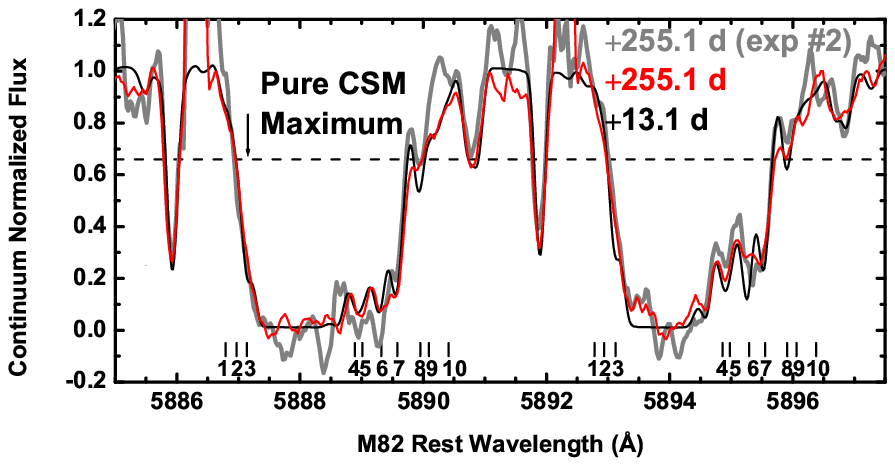}
        \end{minipage}
        \begin{minipage}[]{0.45\textwidth}
                \epsscale{1.0}
                \plotone{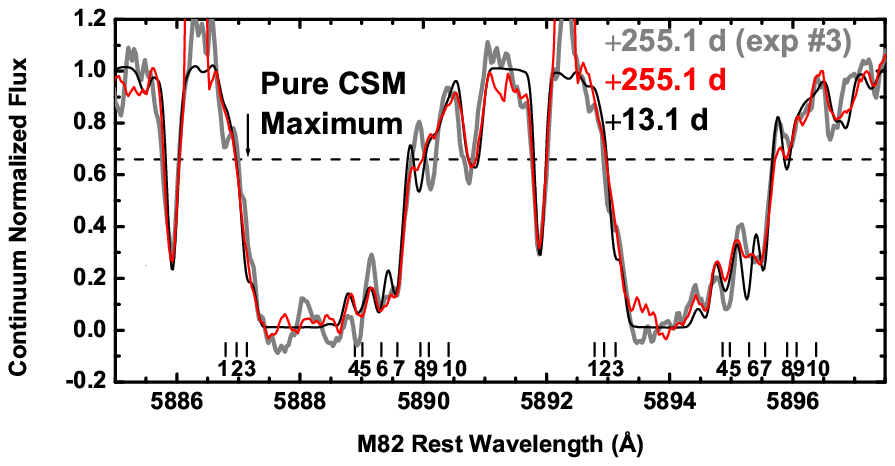}
        \end{minipage}
        \begin{minipage}[]{0.45\textwidth}
                \epsscale{1.0}
                \plotone{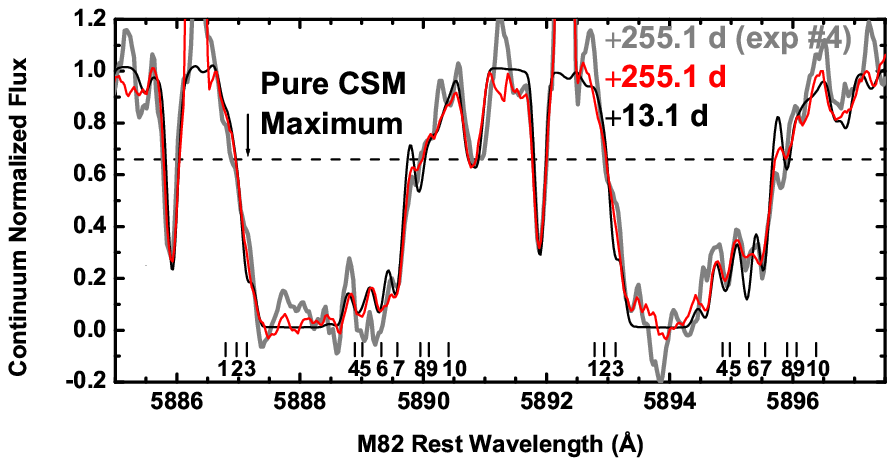}
        \end{minipage}
\end{center}
\caption{(Figure~A3): Spectrum of SN~2014J at $+255.1$~days relative
  to $B$-band maximum brightness with the subtraction of the
  contaminating diffuse component (DC model 1), shown for each
  individual exposure (gray, smoothed with a 7 pixel boxcar filter).
  For comparison, the spectrum extracted from the combined 2D spectrum
  is shown in red (also smoothed with a 7 pixel boxcar filter), as
  well as the $+13.1$~day spectrum in black. \label{figapp3}}
\end{figure*}

\section{Analyses of the \ion{K}{1} Feature}

In this section, we provide a comparison between our $+45.1$~day
spectrum and those from the APF \citep{graham2014}, specifically using
their earliest spectrum taken on $-11$~days and the last one taken on
$+22$~days. To directly compare our spectrum taken by the Subaru
telescope with the APF spectra, it is necessary to deal with a fringe
pattern seen in the APF spectra. In doing this, we make one
assumption -- global patterns of the absorption line spectra are
identical at $+22$ and $+45.1$~days, while allowing the narrow
components at the scale of $\lsim$1~\AA\ to vary as a function of
time. This is possible since the fringing pattern has a characteristic
size of $\gsim$1~\AA\ while the real absorption is at a scale of
$\lsim$1~\AA.

First, the APF spectra are convolved with a Gaussian kernel so that
the spectral resolution is degraded to match the Subaru resolution.
Next, the APF spectrum on $+22$~days was divided by the Subaru
spectrum on $+45.1$~days, resulting in a `fringe pattern function.'
We smoothed this function with a boxcar filter with a width of
$\sim$1~\AA. The result is used as the fringe pattern, and the APF
spectrum of the same epoch was divided by this pattern. The result is
a fringe-corrected spectrum as shown in Figure~12 (blue). While this
assumes that the global pattern did not evolve, if there is a
difference between the two spectra used to generate the fringe pattern
(at $+22$ and $+45.1$~days) at a level narrower than $\sim$1~\AA, one
should be able to detect such features. The same procedure was
performed for the APF spectrum on $-11$~days (gray) where the same APF
spectrum ($-11$~days) was used to create the fringe pattern. As an
additional cross check, we adopted the fringing pattern created with
the APF spectrum on $+22$~days, then corrected the wavelength for the
difference in the heliocentric velocities on $-11$ and $+22$~days, and
finally subtracted this pattern from the spectrum on $-11$~days
(black).  Except for the slight change in the depth of features, which
we attribute to differences in setting the continuum, these two
versions of `fringe-free' $-11$~day spectra are consistent. This
supports the view that the method of subtracting the fringe pattern is robust.

The direct comparison between the Subaru spectrum and the APF spectra
confirms the main conclusions reached by \citet{graham2014}. We do see
the decreasing depths of the \ion{K}{1} components at
$-144$~km~s$^{-1}$ and at $-127$~km~s$^{-1}$ from $-11$~days to
$+22$~days (after correcting for the fringing pattern using the new
Subaru spectrum), while there are no significant changes in the other
\ion{K}{1} components. Comparing the APF
and Subaru spectra, we see no significant evolution from $+20$~days to
$+45.1$~days in the SN~2014J \ion{K}{1} absorption profile.

\end{document}